\documentclass[11pt]{article}
\pdfoutput=1

\usepackage{jheppub}
\usepackage{amsmath}
\usepackage{amssymb}
\usepackage{dcolumn}
\usepackage{bm}
\usepackage{multirow}
\usepackage{blkarray}
\usepackage{slashed}                             
\usepackage{hyperref,graphicx}           
\usepackage{url}
\usepackage{caption}
\usepackage{subcaption}
\usepackage{multirow}
\usepackage{units}
\usepackage{xcolor}
\usepackage{graphicx}
\usepackage{tikz}
\graphicspath{{./Figure/}}

\begin{document}

\title{$b\to s\tau^+\tau^-$ Physics at Future $Z$ Factories}
\author[a]{Lingfeng Li,}
\emailAdd{iaslfli@ust.hk}
\affiliation[a]{Jockey Club Institute for Advanced Study, The Hong Kong University of Science and Technology, Hong Kong S.A.R., P.R.China}
\affiliation[b]{Department of Physics, The Hong Kong University of Science and Technology, Hong Kong S.A.R., P.R.China}
\author[b]{Tao Liu}
\emailAdd{taoliu@ust.hk}
\abstract{$b\to s\tau^+\tau^-$ measurements are highly motivated for addressing lepton-flavor-universality (LFU)-violating puzzles such as $R_{K^{(\ast)}}$ anomalies. The anomalies of $R_{D^{(*)}}$ and $R_{J/\psi}$ further strengthen their necessity and importance, given that the LFU-violating hints from both involve the third-generation leptons directly. $Z$ factories at the future $e^-e^+$ colliders stand at a great position to conduct such measurements because of their relatively high production rates and reconstruction efficiencies for $B$ mesons at the $Z$ pole. To fully explore this potential, we pursue a dedicated sensitivity study in four $b\to s\tau^+\tau^-$ benchmark channels, namely $B^0\to K^{\ast 0} \tau^+ \tau^-$, $B_s\to\phi \tau^+ \tau^-$, $B^+ \to K^+ \tau^+ \tau^-  $ and $B_s \to \tau^+ \tau^-$, at the future $Z$ factories. We develop a fully tracker-based scheme for reconstructing the signal $B$ mesons and introduce a semi-quantitative method for estimating their major backgrounds. The simulations indicate that branching ratios of the first three channels can be measured with a precision $\sim \mathcal O(10^{-7} - 10^{-6})$ and that of $B_s \to \tau^+ \tau^-$ with a precision $\sim \mathcal O(10^{-5})$ at Tera-$Z$. The impacts of luminosity and tracker resolution  on the expected sensitivities are explored. The interpretations of these results in effective field theory are also presented. 
}

\maketitle
\flushbottom
\section{Introduction}
\label{sec:intro}

It has been known for a while that multiple anomalies exist in the measurements of $B$-meson~\footnote{In this paper, ``$B$ mesons'' is defined in a general sense, including $B^{0,\pm}$, $B_c$ and $B_s$, unless otherwise specified (e.g., in the definition of $R_{K^{(\ast)}}$ and $R_{D^{(*)}}$). Similarly, ``$D$ mesons'' is defined to include $D^{0,\pm}$ and $D_s^\pm$.} physics. The first one arises from the $b\to s \ell^+\ell^-$ transitions mediated by flavor changing neutral current (FCNC). It is often termed as $R_{K^{(\ast)}}$ anomaly, with 
\begin{equation}
R_{K^{(\ast)}}\equiv \frac{{\rm BR}(B\to K^{(*)} \mu^+ \mu^-)}{{\rm BR}(B\to K^{(*)}e^+ e^-)}~.
\end{equation}
Here $K^{\ast}$ uniquely refers to $K^\ast(892)^0$, as no other $K$ excited states are of our concern in this work. If lepton flavor universality (LFU) is respected, both $R_K$ and $R_{K^\ast}$ shall be close to one, in the context of Standard Model (SM). However, the measurements indicate that both $R_{K}$ and $R_{K^\ast}$ are lower than this prediction~\cite{Aaij:2017vbb}, with a significance $\sim 2-3 \sigma$. Deviations from the SM predictions also appear in the $b\to c \tau \nu$ and $b\to c l \nu$ transitions which are mediated by flavor-changing charged current (FCCC). Here we omit the neutrino flavor indexes without introducing any ambiguity. One prominent example is 
\begin{equation}
R_{D^{(*)}}\equiv \frac{{\rm BR}(B\to D^{(*)}\tau\nu)}{{\rm BR}(B\to D^{(*)}\ell\nu)}~.
\end{equation}
The anomalies of $\sim 1.3\sigma$ and $\sim2.7\sigma$ have been observed in measuring $R_D$ and $R_{D^*}$, respectively~\cite{Abdesselam:2019dgh}. Another example is 
\begin{equation}
R_{J/\psi}\equiv \frac{{\rm BR}(B_c\to J/\psi \tau\nu)}{{\rm BR}(B_c\to J/\psi \ell\nu)} \ .
\end{equation}
As a counterpart of $R_{D^{(*)}}$ in the $B_c$-meson decays, $R_{J/\psi}$ is measured to be $\sim 2\sigma$ away from its SM prediciton~\cite{Aaij:2017tyk}.
The anomalies of $R_{K^{(\ast)}}$ and $R_{D^{(*)}}$, $R_{J/\psi}$ are summarized in Tab.~\ref{tab:anomalies}. Because of their specific design as a ratio between lepton-flavored quantities, systematic errors are expected to be largely cancelled in measuring these observables. The presence of the said anomalies may indicate the existence of LFU-violating new physics, although more data need to be accumulated before any conclusive statement can be made. 

\begin{table}
\centering
\begin{tabular}{cccc}
\hline 
 & Experimental & SM Prediction & Comments \\ 
\hline 
$R_{K}$  & $0.745^{+0.090}_{ -0.074} \pm0.036$ & $1.00\pm0.01$ \cite{Bordone:2016gaq} & $m_{\ell\ell}^2\in [1.0,6.0]$GeV$^2$, via $B^\pm$. \\ 
$R_{K^\ast}$ & $0.69^{+0.12}_{-0.09}$ & $0.996\pm 0.002$ \cite{Jager:2014rwa} & $m_{\ell\ell}^2\in [1.1,6.0]$GeV$^2$, via $B^0$. \\ 
$R_{D}$ & $0.340\pm 0.030$ & $0.299\pm 0.003$ & $B^0$ and $B^\pm$ combined. \\ 
$R_{D^\ast}$ & $0.295\pm 0.014$ & $0.258\pm 0.005$ & $B^0$ and $B^\pm$ combined. \\ 
$R_{J/\psi}$ & $0.71\pm 0.17\pm 0.18$ & 0.25-0.28~\cite{Aaij:2017tyk} &  \\ 
\hline 
\end{tabular} 
\caption{Summary of the $B$-meson anomalies. The numbers without citations are taken from~\cite{Amhis:2019ckw}.}
\label{tab:anomalies}
\end{table}

Addressing the $R_{K^{(\ast)}}$ anomalies naturally requires extending the FCNC measurements from $b\to s \ell^+\ell^-$ to $b\to s\tau^+\tau^-$. By such measurements, one can probe the LFU-violating phenomena involving the third-generation leptons also. The FCCC anomalies of $R_{D^{(*)}}$ and $R_{J/\psi}$ further strengthen the necessity and significance of the $b\to s\tau^+\tau^-$ measurements, given that the third-generation leptons essentially matter in both of them. In a more general context, this is essential for testing LFU as a fundamental rule in particle physics. Despite these, none of the $b\to s\tau^+\tau^-$ channels have been experimentally observed to date. The measurements of $B_s \to \tau^+ \tau^-$ at LHCb~\cite{Aaij:2017xqt} and $B^+ \to K^+ \tau^+ \tau^-$ at BarBar~\cite{TheBaBar:2016xwe} yield an upper limit of $\mathcal{O}(10^{-3})$ only for their branching ratios. Although these limits can be improved by one to two orders, with an upgrade of LHCb~\cite{Bediaga:2018lhg} and Belle~II~\cite{Kou:2018nap}, they are still far above the SM predictions of $\mathcal{O}(10^{-7})$~\cite{Capdevila:2017iqn,Kamenik:2017ghi}. Instead, the $Z$ factories expected to operate at the next-generation $e^-e^+$ colliders such as CEPC~\cite{CEPCStudyGroup:2018ghi} and FCC-ee~\cite{dEnterria:2016sca} may serve as an ideal platform to measure the $b\to s\tau^+\tau^-$ physics. This strongly motivates a dedicated sensitivity study of $b\to s\tau^+\tau^-$ at the $Z$ pole.

\begin{table}
\centering
\begin{tabular}{cccccc}
\hline 
Channel & Belle~II & LHCb & Giga-$Z$ & Tera-$Z$ & $10 \times$Tera-$Z$   \\ 
\hline 
$B^0$, $\bar{B}^0$ & $5.3\times 10^{10}$ & $\sim 6\times 10^{13}$  & $1.2 \times 10^{8}$  & $1.2 \times 10^{11}$ & $1.2 \times 10^{12}$\\
$B^\pm$ & $5.6\times 10^{10}$ & $\sim 6\times 10^{13}$ & $1.2 \times 10^{8}$  & $1.2 \times 10^{11}$ & $1.2 \times 10^{12}$ \\
$B_s$, $\bar{B}_s$ & $5.7 \times 10^{8}$ & $\sim 2\times 10^{13}$ & $3.2\times 10^{7}$ & $3.2\times 10^{10}$ & $3.2\times 10^{11}$ \\
$B_c^\pm$ & - & $\sim 4 \times 10^{11}$ & $2.2\times 10^5$ & $2.2\times 10^8$ & $2.2\times 10^9$ \\
$\Lambda_b$, $\bar{\Lambda}_b$ & - & $\sim 2\times 10^{13}$ & $1.0\times 10^{7}$ & $1.0\times 10^{10}$ & $1.0\times 10^{11}$ \\
\hline
\end{tabular}
\caption{Number of $b$ hadrons expected to be produced in Belle~II, LHCb and the future $Z$ factories. We assume Belle~II to run at $\Upsilon(4S)$ mode with an integrated luminosity of 50~ab$^{-1}$ and at $\Upsilon(5S)$ with 5~ab$^{-1}$, and estimate the LHCb productions following the $b\bar{b}$ acceptance in~\cite{Albrecht:2017odf}. The production fractions for $B^0/\bar{B}^0$, $B^\pm$, $B_s/\bar{B}_s$ and $\Lambda_b/\bar{\Lambda}_b$ are taken as the average suggested in~\cite{Amhis:2019ckw}.  As for $B_c^\pm$ mesons, we use their production rate at the $Z$ pole presented in~\cite{Zheng:2015ixa} for calculation, with $B_c^\ast$ decays being included, while their production rate at LHCb is taken from a recent measurement~\cite{Aaij:2019ths}. Note, Belle~II will have no statistics on the $B_c^\pm$ and $\Lambda_b/\bar{\Lambda}_b$ productions due to the limitation of energy threshold. } \label{tab:Bnum}
\end{table}

The number of $b$ hadrons expected to be produced in Belle~II, LHCb and the future $Z$ factories is summarized in Tab.~\ref{tab:Bnum}. At Tera-$Z$ as planned for CEPC, the productions of $B^0/\bar{B^0}$ and $B^\pm$ are comparable to those at Belle~II, while the $B_s$/$\bar{B}_s$ productions are nearly two orders more. ILC and FCC-ee are expected to run at the $Z$ pole also, with a plan of Giga-$Z$~\cite{Fujii:2019zll} and upgraded Tera-$Z$ (namely, $10\times$Tera-$Z$)~\cite{Benedikt:2018qee} respectively. Different from these experiments, LHCb produces $b$ hadrons at parton level mainly through QCD processes. The expected productions are typically two to three orders higher than those at Belle~II and Tera-$Z$. 

The future $Z$ factories stand at a great position in measuring the $b\to s \tau^+\tau^-$ physics. They produce $b$ hadrons more boosted, compared to Belle~II. This feature results in a weaker effect of multiple scattering for charged particles such as the ones from the $\tau$-lepton and $D$-meson decays in the tracker, and hence allows their energy/momentum~\cite{Berger:2016vak} and motion direction~\cite{CEPCStudyGroup:2018ghi,Abada:2019zxq} to be measured with higher precision. Moreover, the boosted particles tend to decay with a larger displacement, which may further reduce the uncertainties in reconstructing their decay vertexes. To fully utilize these advantages, we will develop a fully tracker-based scheme for reconstructing the signal $B$ mesons. We will focus on the decay mode of $\tau$ leptons, namely $\tau \to \pi^\pm\pi^\pm\pi^\mp\nu$. Notably, though the LHCb produces $b$ hadrons with a higher rate and more boosted, compared to the other experiments, its data environment is much noisier due to the impacts of QCD effects. This worsens the efficiency and quality of its $B$-meson reconstructions.

\begin{table}[h!]
\centering
\begin{tabular}{ccc}
\hline
Channel &  SM prediction for BR & $q^2\equiv m_{\tau\tau}^2$ (GeV$^2$) \\
\hline
$B^0\to K^{\ast 0} \tau^+ \tau^-$ & $(0.98\pm 0.10)\times 10^{-7}$~\cite{Capdevila:2017iqn} & [15,19]\\
$B_s\to\phi \tau^+ \tau^-$ & $(0.86\pm 0.06)\times 10^{-7}$~\cite{Capdevila:2017iqn} & [15,18.8] \\
$B^+ \to K^+ \tau^+ \tau^-$ & $(1.20\pm 0.12)\times 10^{-7}$~\cite{Capdevila:2017iqn} & [15,22]\\
$B_s \to \tau^+ \tau^-$ & $(7.73\pm 0.49)\times 10^{-7}$~\cite{Kamenik:2017ghi} & -\\
\hline
\end{tabular}
\caption{Four $b\to s\tau^+\tau^-$ benchmark channels. The $B$-meson decay widths are calculated in a phase space selected by $q^2$, where the impacts of the $\psi(2S)$ resonance ($m_{\psi(2S)}=$3.686GeV) on their measurements are suppressed. }
\label{tab:channels}
\end{table}

Below we will pursue a dedicated sensitivity study in four $b\to s \tau^+\tau^-$ benchmark channels (see Tab.~\ref{tab:channels}) at the future $Z$ factories. This study is organized as follows. We develop the scheme of reconstructing the signal $B$ mesons in Section~\ref{sec:pheno}, and discuss their major backgrounds in Section~\ref{sec:pheno2}. Analysis results and their interpretations in effective field theory (EFT) are presented in Section~\ref{sec:result}. We conclude and take an outlook in Section~\ref{sec:conclusion}.

\section{Scheme of Reconstructing the $b\to s\tau^+\tau^-$ Events}
\label{sec:pheno}

In this study, the $e^+e^-\to Z \to b \bar b$ events and their shower are simulated using Pythia8~\cite{Sjostrand:2007gs}. We decay $B$ mesons exclusively and their intermediate particles ($\tau$ leptons, $D$ mesons, etc.) inclusively. The decays of $\tau^\pm \to\pi^{\pm}\pi^{\pm}\pi^{\mp}\nu$ and $\tau^\pm \to \pi^\pm\pi^\pm\pi^\mp \pi^0 \nu$ are modeled respectively by the CLEO~\cite{Asner:1999kj}, with an intermediate state of $a_1(1260)^\pm \to \rho(770)^0\pi^\pm$ mostly~\cite{Asner:1999kj,Ilten:2012zb}, and the Novobrisk~\cite{Bondar:2002mw}, with an intermediate state of $\pi^\pm\omega(782)$ instead. The signal events are generated only for the $q^2\equiv m_{\tau\tau}^2$ windows defined in Tab.~\ref{tab:channels}. The detector effects are simulated using Delphes3~\cite{deFavereau:2013fsa}, with a CEPC-detector template~\cite{Chen:2017yel} being applied. 

\begin{figure}[h!]
\centering
\includegraphics[height=4.8 cm]{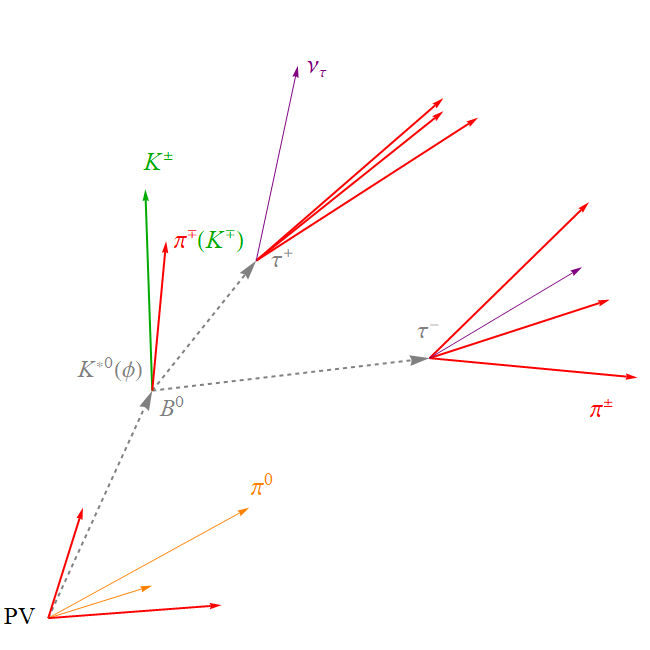}
\includegraphics[height=4.8 cm]{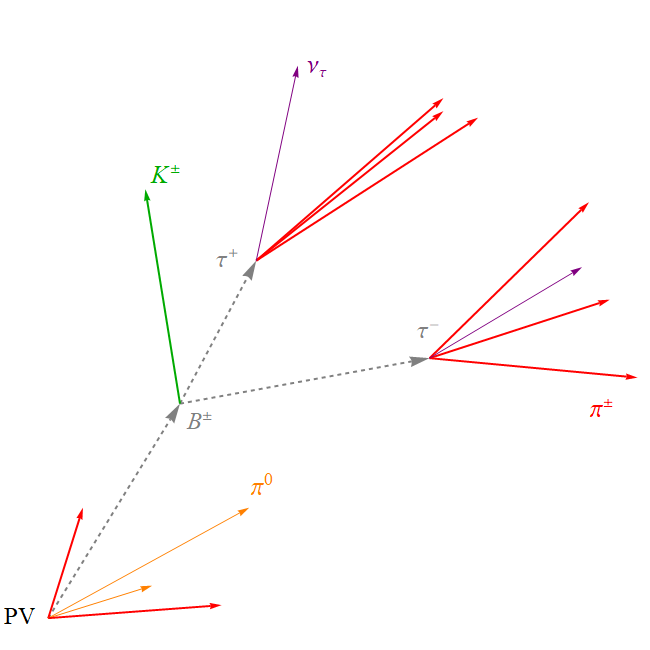}
\includegraphics[height=4.8 cm]{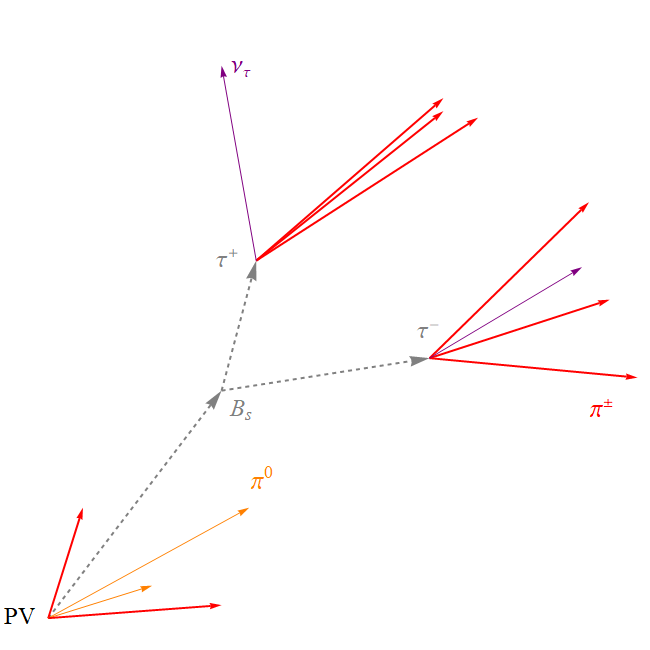}
\caption{Schematic pictures of the $B^0\to K^{\ast 0} (\phi) \tau^+ \tau^-$ (left), $B^+ \to K^+ \tau^+\tau^-$ (middle) and $B_s\to\tau^+ \tau^-$ (right) events. The dashed arrows represent spatial displacement of $B$ mesons and $\tau$ leptons between their production and decay vertexes ($\vec{V}$), and the solid ones denote three-momenta ($\vec{p}$) of their decay products or accompanying particles.  }
\label{fig:schematic}
\end{figure}

As discussed above, our analysis will focus on the four benchmark channels of $b \to s \tau^+\tau^-$ listed in Tab.~\ref{tab:channels} with $\tau^\pm \to\pi^{\pm}\pi^{\pm}\pi^{\mp}\nu$. These events have no neutral particles except neutrinos in their final states. Then the yet-to-be-determined neutrino momenta leave six d.o.f. to fix for the $B$-meson reconstruction. We demonstrate the topologies of these four classes of events in Fig.~\ref{fig:schematic}. For $B^0\to K^{\ast 0} (\phi) \tau^+ \tau^-$, the decay vertex of $B^0$ can be fully reconstructed due to the prompt decay of $K^{\ast 0} \to K^+ \pi^- \ (\phi \to K^+ K^-)$.  But for $B^+ \to K^+ \tau^+\tau^-$, the $B^+$ decay vertex can be constrained to be along $\vec{V}_{K^+}$ only. The story is even worse for $B_s\to\tau^+ \tau^-$, where the $B_s$ decay vertex is invisible to the detector at all. In view of the differences, below we will develop dedicated strategies for reconstructing these benchmark $B$-meson events.

Let us start with the measurements of $B^0\to K^{\ast 0} \tau^+ \tau^-$ and $B_s\to\phi \tau^+ \tau^-$. A preselection of these events is applied to ensure their successful reconstruction.  We first require for each $B^0\to K^{\ast 0} \tau^+ \tau^-$ event a $K^\pm \pi^\mp$ vertex with its invariant mass being in $[m_{K^{\ast 0}}- 0.1,m_{K^{\ast 0}}+ 0.1]$GeV, and for each $B_s\to\phi \tau^+ \tau^-$ event a $\phi\to K^+ K^-$ vertex with its invariant mass being in $[m_\phi- 0.02,m_\phi+ 0.02]$GeV. Here the allowed mass ranges for these vertexes are set by the width of these mesons and the detector resolution. The distance of these vertexes to the PV needs to be greater than 0.5mm. Moreover, we require two candidate $\tau\to\pi^\pm\pi^\pm\pi^\mp\nu$ vertexes for each event, with their distance to the PV being larger than $0.5$mm and the geometric angle $\Delta \Omega (\vec{p}_{3\pi,i},\vec{p}_{K^{\ast 0}(\phi)})$ between the reconstructed $\vec{p}_{3\pi,i}$ and $\vec{p}_{K^{\ast 0} (\phi)}$ being smaller than one. As a physical requirement, the energy of each $B$ meson should be smaller than $m_Z/2$. Finally, to match the theoretical predictions listed in Tab.~\ref{tab:channels}, we require the reconstructed $m_{\tau\tau}^2$ to be in $[15,19(18.8)]$GeV$^2$ for the $B^0\to K^{\ast 0} \tau^+ \tau^-(B_s\to\phi \tau^+ \tau^-)$ events. 

\begin{figure}[h!]
\centering
\includegraphics[height=5. cm]{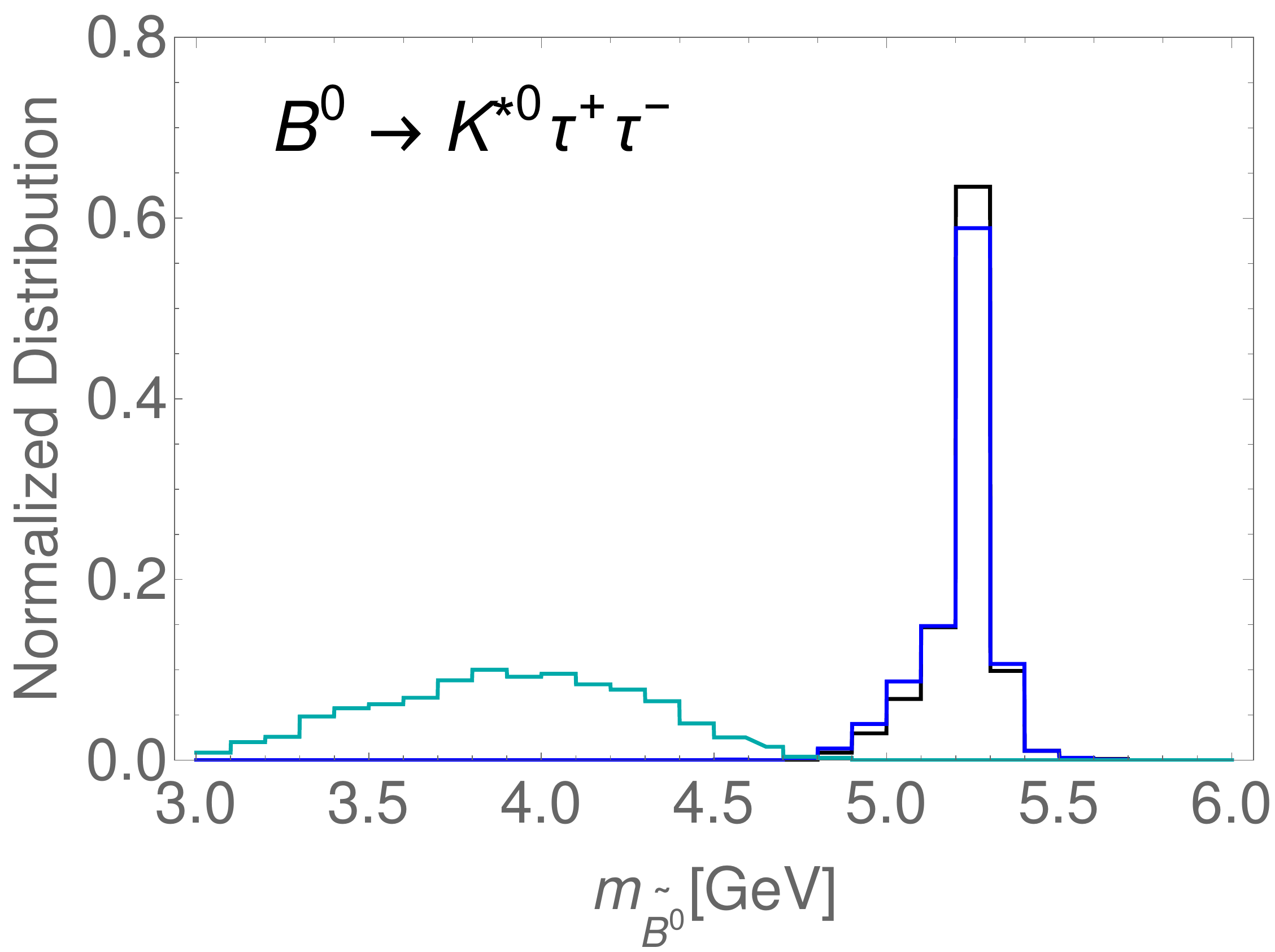} \ \ \ \ \ \ 
\includegraphics[height=5. cm]{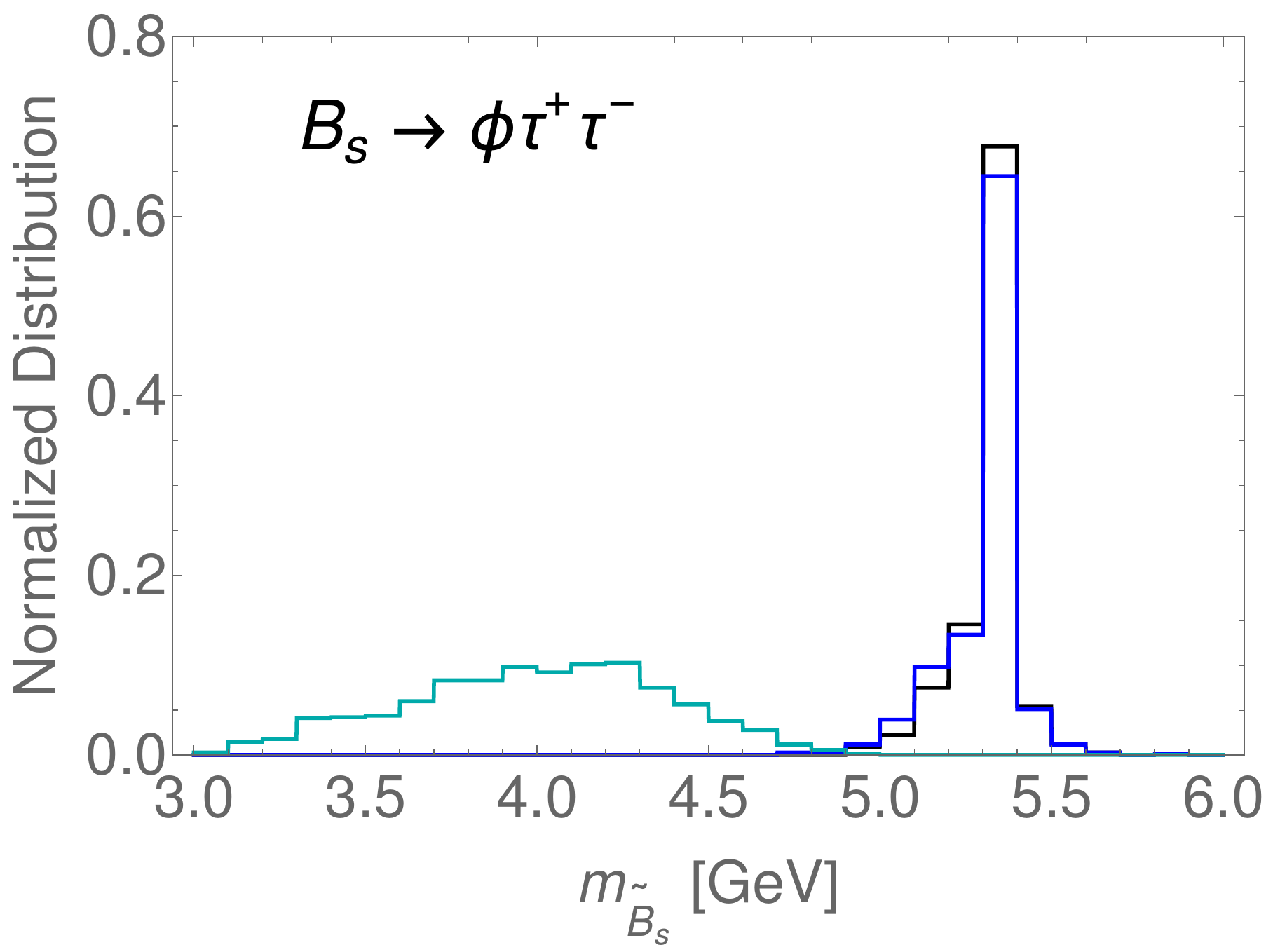}
\includegraphics[height=5. cm]{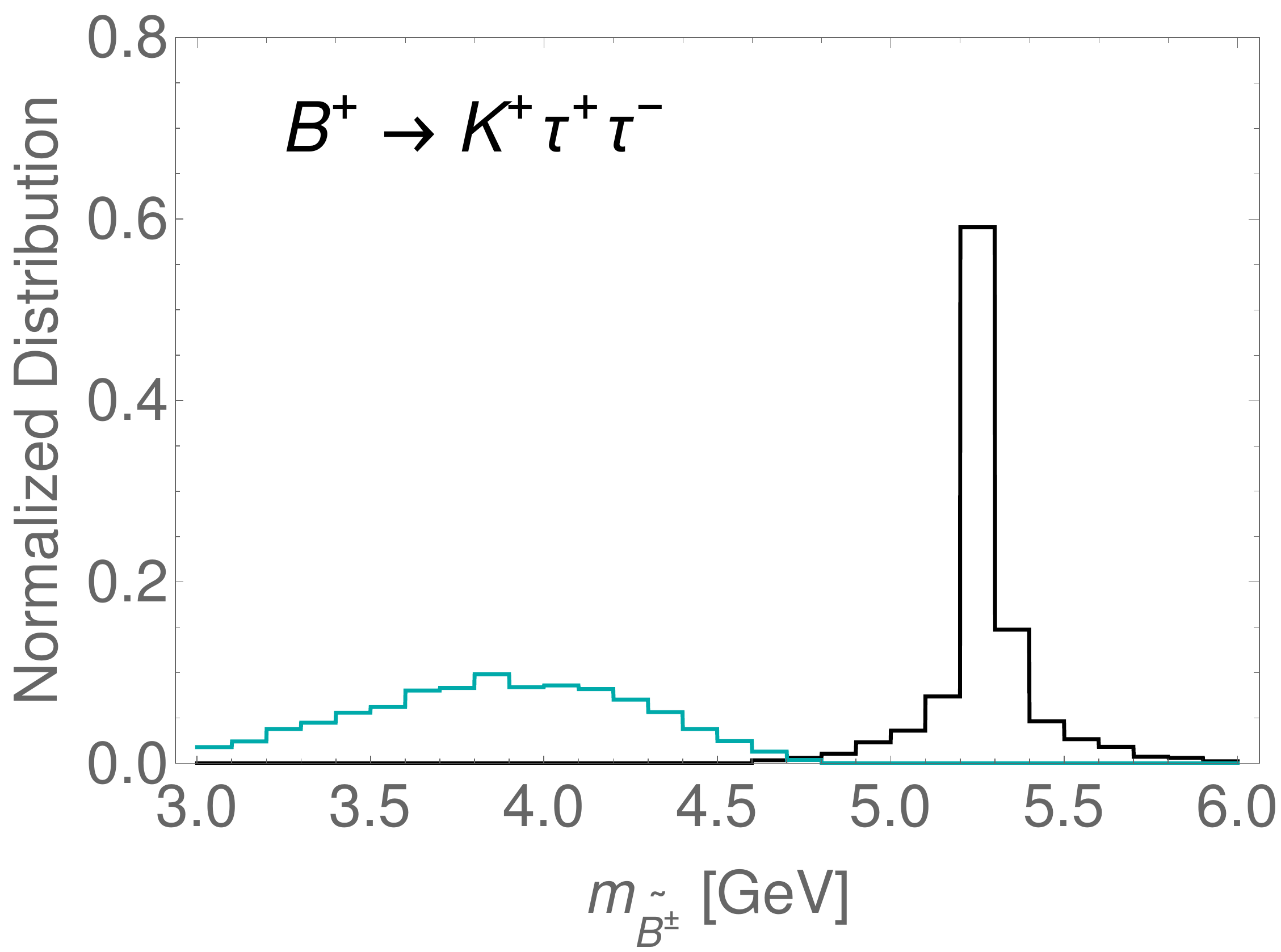} \ \ \ \ \ \ 
\includegraphics[height=5. cm]{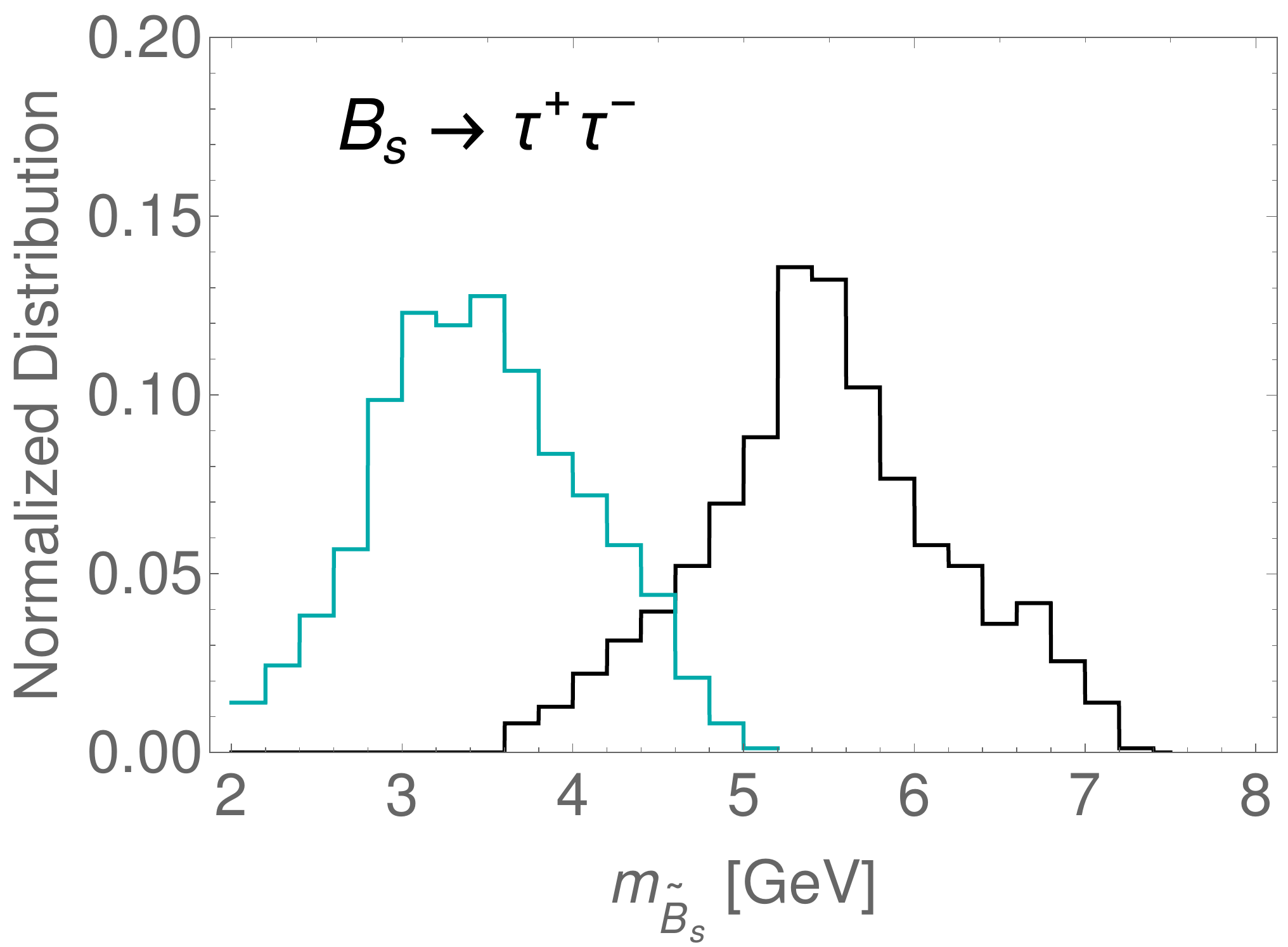}
\caption{Normalized mass distributions of the reconstructed $B$ mesons. The cyan curves, representing the invariant mass of the charged particles from the $B$-meson decays, serve as a reference. The blue curves in the upper panels result from an intermediate reconstruction, $i.e.$, solving the equations in Eq.~(\ref{eq:fullkinematics1}). The black curves are obtained from the full reconstruction by minimizing the reconstruction error $\epsilon$.}.
\label{fig:rec}
\end{figure}

The $B^0\to K^{\ast 0} \tau^+ \tau^-$ and $B_s\to\phi \tau^+ \tau^-$ events are reconstructed in the same way. For simplicity, we will use $B^0\to K^{\ast 0} \tau^+ \tau^-$ to demonstrate this strategy. The suggested reconstruction relies on three relations:  
\begin{equation}
\label{eq:fullkinematics1}
\vec{p}_{B^0} \times \vec{V}_{B^0}=0~ \Rightarrow (\vec{p}_{K^{\ast 0}} + \sum_{i=1,2} \vec{p}_{\tau,i} ) \times \vec{V}_{B^0}=0~,  \ \ \ \  \vec{p}_{\tau,i} \times \vec{V}_{\tau,i}=0~(i=1,2)\ .
\end{equation}
Here ``$i$'' runs over the two $\tau$ leptons. These relations manifest that $B^0$ meson and $\tau$ leptons displace along their respective momentum directions. Because the momentum magnitudes can be scaled out, these relations provide totally six independent kinematic conditions and hence are able to determine the neutrino momenta as ($i\neq j$) 
\begin{equation}
\vec{p}_{\nu,i} = \frac{-\vec{p}_{K^{\ast 0}}\times \vec{V}_{B^0}\cdot \vec{V}_{\tau,j}}{\vec{V}_{\tau,i}\times \vec{V}_{B^0}\cdot \vec{V}_{\tau,j}} \vec{V}_{\tau,i} - \vec{p}_{3\pi,i}~.
\label{eq:fullkinematics3}
\end{equation}
The normalized mass distributions of the reconstructed $B$ mesons are shown in Fig.~\ref{fig:rec}. Compared to the reference, namely the invariant mass of the charged particles from the $B$-meson decays, one can see that the quality of reconstruction is greatly improved, yielding a sharp peak around $m_{B^0}$. 

The reconstruction can be further improved using the on-shell conditions of $\tau$ leptons. For that case, we allow a variation to the reconstructed neutrino momentum, namely $\delta \vec{p}_{\nu,i}$, along its reconstructed direction. Then the favored value of $\delta \vec{p}_{\nu,i}$ can be determined by minimizing the reconstruction error
\begin{equation}
\epsilon (\delta \vec{p}_{\nu,i}) =  \sum_{i=1,2} \frac{[m_{\tilde \tau,i} - m_\tau]^2}{m_\tau^2} + \sum_{i=1,2} \frac{|\delta \vec{p}_{\nu,i}|^2}{|\vec{p}_{\nu,i}|^2} ~.
\label{eq:reconstructionerror1}
\end{equation}
Here (and below) the symbol of tilde is applied to denote the reconstructed ``resonance''. The second term is introduced as a regulator of the loss caused by the variation of $\vec{p}_{\nu,i}$. The $B$ mass distributions based on this method are also shown in Fig.~\ref{fig:rec}. Indeed, the reconstruction gets slightly improved, compared to the method based on Eq.~(\ref{eq:fullkinematics3}) only, eventually resulting in a sharper peak around $m_{B}$ in both upper-left and upper-right panels. Below we will take this optimized method for the $B \to K^* \tau^+\tau^-$ and $B_s \to \phi\tau^+\tau^-$ analyses.

The preselection of the $B^+ \to K^+ \tau^+ \tau^-$ events is largely similar to those for $B^0\to K^{\ast 0}  \tau^+\tau^-$ and $B_s\to \phi  \tau^+\tau^-$. We require each event to have one candidate $K^\pm$ track, with its minimal distance to the PV being greater than $0.1$mm, and two candidate $\tau$ vertexes, with  their distances to the PV being greater than $0.5$mm and the geometric angle $\Delta \Omega(\vec{p}_{3\pi,i},\vec{p}_{K^\pm})$ between the reconstructed $\vec{p}_{3\pi,i}$ and $\vec{p}_{K^\pm}$ being smaller than one. We also require the energy of the reconstructed $B$ meson to be smaller than $m_Z/2$. Finally, the reconstructed $m_{\tau\tau}^2$ is required to be in $[15,22]$GeV$^2$, to match the theoretical prediction listed in Tab.~\ref{tab:channels}.

For the $B^+ \to K^+ \tau^+ \tau^-$ events, the $B^\pm$ vertex can be constrained to along the $K^\pm$ track only, yielding 
\begin{equation}
\vec{V}_{B^\pm} = \vec{V}_{B^\pm}^{\rm ini} + \alpha \vec{n}_{K^\pm}~.
\end{equation}
Here $\vec{V}_{B^\pm}^{\rm ini}$ is a vector directed from the PV to its closest point on the $K^\pm$ track, $\vec{n}_{K^\pm}$ is an unit vector along this track, and $\alpha$ is a free parameter to measure the shift of $B^\pm$ in that direction. As the $B^\pm$ version of Eq.~(\ref{eq:fullkinematics3}) can be satisfied with arbitrary $\alpha$ values, $\alpha$ is left unconstrained by the counterpart of Eq.~(\ref{eq:fullkinematics1}). We will find the favored $\alpha$ value then by minimizing the reconstruction error
\begin{equation}
\epsilon (\alpha) =   \sum_{i=1,2} \frac{[m_{\tilde \tau,i} - m_\tau]^2}{m_\tau^2}  ~.
\end{equation}
Again, the on-shell conditions of $\tau$ leptons are applied here. From Fig.~\ref{fig:rec}, one can see that the $B^+$ events are well-reconstructed, yielding a sharp peak around $m_{B^+}$ also.

The $B_s\to \tau^+\tau^-$ preselection is relatively simple. It is implemented by requireing each event to have two candidate $\tau$ vertexes, with their distances to the PV being greater than $0.5$mm and the geometric angle $\Delta\Omega(\vec{p}_{3\pi,1},\vec{p}_{3\pi,2})$ between $\vec{p}_{3\pi,1}$ and $\vec{p}_{3\pi,2}$ being smaller than one, and moreover, the energy of the reconstructed $B_s$ meson to be less than $m_Z/2$.

But, the reconstruction of $B_s\to\tau^+\tau^-$ is more involved, compared to the other three cases. The $B_s$ vertex can not be located at all and hence brings in three extra d.o.f., namely $\Vec{V}_{B_s}$. Even if we combine the self-consistency relations
\begin{equation}
\Vec{V}_{B_s}\times \vec{p}_{B_s}=\Vec{V}_{B_s}\times (\vec{p}_{\tau,1}+\vec{p}_{\tau,2})=0 ~,~ \vec{V}_{\tau,i}\times \vec{p}_{\tau,i}=0~(i=1,2)
\label{eq:Bsconsisitancy}
\end{equation}
and the on-shell conditions of $\tau$ leptons, there is still one parameter left unfixed~\footnote{In this study, the $B$-meson mass will be used as a discriminator to distinguish between the signal and its backgrounds, instead of an on-shell condition for the event reconstruction.}. In order to address this, we first formally rewrite Eq.~(\ref{eq:Bsconsisitancy}) as
\begin{equation}
{\bm V} = {\bm H} {\bm P}~;~{\bm V} = \left(\begin{matrix} \vec{V}_{\tau,1}+\vec{V}_{B_s}\\ \vec{V}_{\tau,2}+\vec{V}_{B_s}\end{matrix}\right)~,~{\bm H} = \left(\begin{matrix}h_{\tau,1}+h_{B_s} &h_{B_s} \\ h_{B_s} &h_{\tau,2}+h_{B_s} \end{matrix}\right)~,~{\bm P} = \left(\begin{matrix} \vec{p}_{\tau,1}\\ \vec{p}_{\tau,2}\end{matrix}\right),
\label{eq:selfconsistency}
\end{equation}
with $h_i \equiv \frac{\vec{V}_i}{\vec{p}_i} \ (i=\tau_1,\tau_2,B_s)$. Here the two components of ${\bm V}$ can be measured directly. The three extra d.o.f of $\vec{V}_{B_s}$ are then transformed into the new $h_i$ parameters. Among the nine free parameters to reconstruct, seven can be solved in terms of $\frac{\vec{p}_{\nu,i}\cdot \vec{p}_{3\pi,i} }{|\vec{p}_{3\pi,i}|}$, by combining the two on-shell conditions of $\tau$ leptons and five of the six self-consistency equations in Eq.~(\ref{eq:selfconsistency}). The reconstruction eventually can be achieved by minimizing the reconstruction error 
\begin{equation}
\epsilon \left(\frac{\vec{p}_{\nu,i}\cdot \vec{p}_{3\pi,i} }{|\vec{p}_{3\pi,i}|}\right) =  \sum_{i=1,2} \frac{[m_{\tilde \tau,i} - m_\tau]^2}{m_\tau^2} + \sum_{i,j =1,2}\frac{|({\bm V}-{\bm H}{\bm P})^T({\bm V}-{\bm H}{\bm P})|_{ij}}{|{\bm V^T \bm V}|_{ij} }~.
\label{eq:ErrorBs}
\end{equation}
A reconstruction error similar to this one can be also found in~\cite{Morda:2015lit}. Though the $B_s$-meson reconstruction is possible in this setup, its quality suffers from the insufficiency of the constraints. As shown in Fig.~\ref{fig:rec}, the mass distribution for the reconstructed $B_s$ mesons is much broader than those of the other three cases.

Notably, we have assumed the tracker to be perfect in spatial resolution so far in the discussions on the event reconstruction. As demonstrated in Fig.~\ref{fig:rec}, the invariant mass of the signal $B$ mesons can be well reconstructed (except the $B_s$ mesons of $B_s \to \tau^+\tau^-$) in this ideal case, yielding a sharp peak in its distributions around the physical value. However, we will see later that the major backgrounds for these analyses arise from the $B$-meson decays. The reconstructed invariant mass for the background events actually is not far from that of the signal ones. The finite resolution of the tracker thus may significantly impact the sensitivities that one can reach at the future $Z$ factories and hence can not be neglected. We will study this in detail in Section~\ref{sec:result}.

\section{Major Backgrounds}
\label{sec:pheno2}

\subsection{Exclusive $B$-Meson Decays}
\label{ssec:background}

In our scheme, di-$\tau$ identification plays a crucial role in reconstructing the $b\to s \tau^+ \tau^-$ events. This is expected to be achieved by requiring two $\pi^\pm\pi^\pm\pi^\mp$ vertexes displaced from the PV or the $B$-meson vertex. In the SM, $D_s^\pm$ and $D^\pm$ mesons are known to have a mass and lifetime comparable to those of $\tau$ leptons. They can decay to a $\pi^\pm\pi^\pm\pi^\mp+X$ state with non-negligible branching ratios, as shown in Tab.~\ref{tab:ddecay}. Here $X$ represents neutral particles. Thus we expect the major backgrounds for the $b\to s \tau^+\tau^-$ measurements to arise from the Cabibbo-favored $b\to c+X$ processes (as also noticed in~\cite{Kamenik:2017ghi,Aaij:2017xqt}), namely $b\to c\bar{c}s$, $b\to c\tau \nu$ and $b\to c\bar{u}d$, where either one or both $\tau$ leptons in the signal are faked by charged $D$ meson(s) with the said decays~\footnote{In the analysis, we also take into account the $b\to c+X$ events with excited $D$ meson(s), namely $D^{\ast\pm}$ and $D_s^{\ast\pm}$. In these events, the excited $D$ meson decays back to its ground-state $D$ meson and a neutral pion or photons. The $\tau$ lepton then can be faked by the ground-state $D$ meson.}. For the measurement of $B_s\to \tau^+\tau^-$, $b\to c\bar{u}d$ is especially important, where the event reconstruction is less constrained and hence the chance for one $\tau$ lepton to be faked by three charged pions from the $B$-meson decay directly can not be neglected.

\begin{table}[h]
\centering
\begin{tabular}{c|ccc}
\hline
                 & Properties & Decay Mode & BR \\
\hline                  
\multirow{2}{1.cm}{$\tau^\pm$} & \multirow{2}{3cm}{$m=1.777$GeV $L=87.0\mu$m} & $\pi^\pm\pi^\pm\pi^\mp\nu $ &$ 9.3\%$ \\
                  & & $\pi^\pm\pi^\pm\pi^\mp\pi^0 \nu $   & $4.6\%$ \\

\hline                  
\multirow{5}{1.cm}{$D_s^\pm$} & \multirow{5}{3cm}{$m=1.968$GeV $L=151\mu$m} & $\tau^\pm \nu $ &$5.5\%$ \\
                &  & $\pi^\pm\pi^\pm\pi^\mp $   & $1.1\%$ \\
                &  & $\pi^\pm\pi^\pm\pi^\mp \pi^0  $ & $0.6\%$ \\
                &  & $\pi^\pm\pi^\pm\pi^\mp 2\pi^0 $ & $ 4.6\%$ \\
                &  & $\pi^\pm\pi^\pm\pi^\mp K_S^0 $ & $0.3\%$ \\
                &  & $\pi^\pm\pi^\pm\pi^\mp \phi $ & $1.2\%$ \\
\hline                  
\multirow{3}{1.cm}{$D^\pm$} & \multirow{3}{3cm}{$m=1.870$GeV $L=311\mu$m} & $ \tau^\pm \nu$ & $< 0.12\%$ \\
                &  & $\pi^\pm\pi^\pm\pi^\mp $ & $0.31\%$  \\
                &  & $\pi^\pm\pi^\pm\pi^\mp\pi^0 $ & $1.1\%$ \\
                &  & $\pi^\pm\pi^\pm\pi^\mp K_S^0 $ & $3.0\%$ \\
\hline                  
\end{tabular}
\caption{Invariant mass and mean decay length ($L$) of $\tau$ leptons and charged $D$ mesons~\cite{Tanabashi:2018oca}. Some decay modes relevant to this study and their branching ratios are also presented.}
\label{tab:ddecay}
\end{table}

Notably, the misidentification between $\pi^\pm$ and $K^\pm$, if its rate is high, will cause extra complexity for evaluating the backgrounds. First, $D$ mesons with a decay mode such as $D^\pm \to K^\mp\pi^\pm\pi^\pm\pi^0$ can fake $\tau$ lepton with a large chance. Their contribution to the backgrounds thus may not be negligibly small. Second, the $K^{\ast 0}\to K^+ \pi^-$ and $\phi\to K^+K^-$ vertexes can be mutually faked, such that $B^0\to K^{\ast 0} \tau^+ \tau^-$ and $B^0\to \phi \tau^+ \tau^-$ events will serve as the backgrounds mutually in their measurements. Also, for the $B^+\to K^+ \tau^+ \tau^-$ measurement, extra backgrounds need to be taken into account where $K^+$ is mimicked by $\pi^+$. However, $K^\pm$ and $\pi^\pm$ could be well-separated in the tracking system of the future $Z$ factories~\footnote{We would greatly thank Franco Grancagnolo and Manqi Ruan for communications regarding this point.}. For $|\vec{p}|\gtrsim 2$GeV, the separation~\cite{Lippmann:2011bb} between $\pi^\pm$ and $K^\pm$ can be more than 4$\sigma$ in terms of the number of primary ionization clusters formed in the drift chamber~\cite{CEPCStudyGroup:2018ghi,Abada:2019zxq}. Below this threshold, $K^\pm$ moves much slower than $\pi^\pm$, because of its bigger mass. The difference of arrival time between $K^\pm$ and $\pi^\pm$ at the first electromagnetic calorimeter layer can be well-captured by a detector with a time resolution $\lesssim\mathcal{O}(100)$~ps~\cite{Sadrozinski:2017qpv}. Actually, a detector with such a time resolution has already been part of the CMS upgrade plan~\cite{Collaboration:2296612}. Therefore, we expect the backgrounds stemming from the misidentification between $\pi^\pm$ and $K^\pm$ to be well-controlled at the future $Z$ factories, and hence will not consider them below. 

Yet, we are still confronted with evaluating the branching ratios of the $B$-meson decays said above and hence the rates of the major backgrounds. So far, only a small fraction of them have been experimentally measured or theoretically calculated. To address this difficulty, we will take the following strategy. We introduce first  some $B$-meson decay mode which has been well-measured as a reference, then evaluate the suppression factor of the width of the background $B$-meson decay relative to the reference one, and at last calculate its branching ratio by scaling, assuming the decay widths of $B^0$, $B^\pm$ and $B_s$ to be the same~\footnote{In the SM, the total decay widths of $B^0$, $B^\pm$ and $B_s$ are different by less than $10\%$~\cite{Tanabashi:2018oca}. We will tolerate the inaccuracy caused by neglecting this difference in our analysis, given that the uncertainties brought in by the said strategy for evaluating the backgrounds could be more significant.}. This strategy is semi-quantitative. But we expect the improvements for the sensitivity analysis to be straightforward in this framework whenever new inputs become available. Below is a summary of the main effects which potentially contribute to the suppression factor, as reviewed in~\cite{Bevan:2014iga}.
\begin{itemize}
\item Suppression caused by color factor. At tree level, the Feynman diagrams of the $b\to c\bar{c}s$ and $b\to c\bar{u}d$ decays can be either external or internal in terms of $W$ emission. It depends on the hadronization topology of the quarks from $W$ decay. For an internal $W$-emission diagram, after hadronization the $W$ decay products become color-connected with the remaining $B$-meson decay products. The color factor in this case is three times smaller than that of the external-$W$ emission.

\item Suppression arising from the phase-space limitation. The phase space available for the $B$-meson decays is approximately set by the mass difference between the $B$ meson and its decay products. For the decay mode with extra heavy particles being produced, its width will be suppressed by the relative smallness of the phase space available for decay. Two cases are relevant here. In the first one, the phase space is squeezed by the production of an $s\bar{s}$ pair out of the vacuum. Another one is related to the perturbative $\tau$ production (e.g., $b\to c\tau\nu$ vs. $b\to c\ell \nu$), where the phase space is limited by its mass.

\end{itemize}
Moreover, we introduce a rule of analogy, mainly in terms of decay topology and isospin symmetry of $B$ mesons, for assisting our estimation of the suppression factor. This will allow us to use the existing experimental measurements to estimate the suppression factor as an analogue in some contexts, and hence strengthens the applicability and accuracy of this strategy. 

In Tab.~\ref{tab:B0Ksbackground}-\ref{tab:Ditaubackground}, the branching ratios presented with a reference are from experimental measurements, while the others are based on our estimation using the suggested strategy. As a test of this strategy, let us consider $B^+ \to K^+ D^{(\ast)+} D^{(\ast)-}$ in Tab.~\ref{tab:BcKcbackground}, using $B^+ \to K^0 \bar{D}^{(\ast)0}D^{(\ast)+}$ as a reference. According to the rule of analogy, we have 
\begin{equation}
\frac{\text{BR}(B^+ \to K^+ D^{(\ast)+} D^{(\ast)-})}{\text{BR}(B^+ \to K^0\bar{D}^{(\ast)0}D^{(\ast)+})} = \frac{\text{BR}(B^0 \to K^0\bar{D}^{(\ast)0} D^{(\ast)0} )}{\text{BR}(  B^0\to K^+  D^{(\ast)-}D^{(\ast)0})} \sim 0.21. 
\label{test}
\end{equation}
Here $B^+ \to K^+ D^{(\ast)+} D^{(\ast)-}$ and $B^0 \to K^0\bar{D}^{(\ast)0} D^{(\ast)0}$ are color-suppressed, while $B^+ \to K^0 \bar{D}^{(\ast)0}D^{(\ast)+}$ and $B^0\to K^+  D^{(\ast)-}D^{(\ast)0}$ are not. All of their branching ratios have been measured. The ratio of 0.21 in Eq.~(\ref{test}) is determined by the existing measurements for the two $B^0$ decay channels~\cite{Amhis:2019ckw}. Then given BR($B^+ \to K^0 \bar{D}^{(\ast)0}D^{(\ast)+}) = 1.7\%$~\cite{Amhis:2019ckw}, we find $\text{BR}(B^+ \to K^+ D^{(\ast)+} D^{(\ast)-}) \sim 3.5\times 10^{-3}$, about $25\%$ bigger than its measured value, namely $2.8\times 10^{-3}$~\cite{Tanabashi:2018oca}. We can also estimate the $R_{D^{(\ast)}}$ and $R_{J/\psi}$ values using this strategy. The estimations based on the suppression caused by the $\tau$ production (see details below for this method) turn out to deviate from the SM predictions shown in Tab.~\ref{tab:anomalies} by less than $10\%$. These tests justify this strategy to a good extent.

\begin{table}[h]
\centering
\begin{tabular}{c|ccccc}
\hline
Type & Channel & Color & $s\bar{s}$ & $\tau$ & BR  \\
\hline 
\hline 
\multirow{4}{1.5cm}{$b\to c\bar{c}s$} 
&$B^0\to K^{\ast 0} D^{(\ast)+} D^{(\ast)-}$ & & & & $1.2\times 10^{-2}$\\
&$B_s\to K^{\ast 0} D^{(\ast)+} D_s^{(\ast)-}$ & & & & $1.2\times 10^{-2}$ \\ 
&$B_s\to \bar{K}^{\ast 0} D_s^{(\ast)+} D^{(\ast)-}$ & & & & $1.2\times 10^{-2}$ \\
&$B^0\to K^{\ast 0} D_s^{(\ast)+} D_s^{(\ast)-}$ & & \checkmark & & $ 1.6 \times 10^{-3}$ \\
\hline 
\multirow{2}{1.5cm}{$b\to c\tau\nu$} & $B^0\to K^{\ast 0} D_s^{(\ast)-}\tau^+ \nu$ & &\checkmark & \checkmark & $ 3.0\times 10^{-5}$ \\
&$B_s\to \bar{K}^{\ast 0} D^{(\ast)-}\tau^+\nu$ & & & \checkmark & $ 4.6 \times 10^{-4}$ \\
\hline
\end{tabular} 
\caption{Major backgrounds for the $B^0\to K^{\ast 0} \tau^+ \tau^-$ measurement. ``Color'', ``$s\bar{s}$'' and ``$\tau$'' represent the three effects suppressing the $B$-meson decays which are discussed in the text.}
\label{tab:B0Ksbackground}
\end{table}

The major backgrounds for the $B^0\to K^{\ast 0} \tau^+ \tau^-$ measurement are summarized in Tab.~\ref{tab:B0Ksbackground}. As a start, we consider $B^0\to K^{\ast 0} D^{(\ast)+} D^{(\ast)-}$ and take $B^0\to K^0 D^{(\ast)+} D^{(\ast)-}$ as a reference. Both decay modes are color-unsuppressed. Then following the rule of analogy, we have
\begin{equation}
\frac{\text{BR}(B^0\to K^{\ast 0} D^{(\ast)+} D^{(\ast)-} )}{\text{BR}(B^0\to  K^0 D^{(\ast)+} D^{(\ast)-})} = \frac{\text{BR}(B^0\to K^{\ast 0} D^0 \bar{D}^{0}  )}{\text{BR}(B^0\to K^0 D^0 \bar{D}^{0}   )} \sim 0.81~.
\label{eq:BRrelation2}
\end{equation}
Here BR($B^0\to K^0 D^0 \bar{D}^{0})$ has been measured to be $2.7\times 10^{-4}$~\cite{delAmoSanchez:2010pg} and BR($B^0\to K^0 D^0 \bar{D}^{0})\sim   2.2\times 10^{-4}$ is extracted out from the inclusive $B^0\to K^+ \pi^- D^0 \bar{D}^{0}$ measurement with $m_{K^+\pi^-}\in[0.8,1.0]$GeV~\cite{Aaij:2020qup}. Given BR$(B^0 \to  K^{ 0} D^{(\ast)+} D^{(\ast)-}) =1.5\%$~\cite{Amhis:2019ckw}, we immediately find BR$(B^0\to K^{\ast 0} D^{(\ast)+} D^{(\ast)-}) \sim 1.2\%$. BR($B_s\to K^{\ast 0} D^{(\ast)+} D_s^{(\ast)-}$) and BR($B_s\to \bar{K}^{\ast 0} D_s^{(\ast)+} D^{(\ast)-}$) can be estimated similarly. Since these two decay modes share the same topology with $B^0\to K^{\ast 0} D^{(\ast)+} D^{(\ast)-}$ and meanwhile do not suffer the three suppression effects said above, we simply assume their branching ratios to the same as BR$B^0\to K^{\ast 0} D^{(\ast)+} D^{(\ast)-}$. 

The remaining $b\to c\bar{c}s$ background, namely $B^0\to K^{\ast 0} D_s^{(\ast)+} D_s^{(\ast)-}$, is suppressed by the $s\bar{s}$ production. To estimate its branching ratio, we take $B^0\to K^{\ast 0} D^{(\ast)+} D^{(\ast)-}$, a mode not suppressed by the $s\bar{s}$ production, as the reference and assume 
\begin{equation}
\frac{\text{BR}(B^0\to K^{\ast 0} D_s^{(\ast)+} D_s^{(\ast)-})}{\text{BR}(B^0\to K^{\ast 0} D^{(\ast)+} D^{(\ast)-} )} = \frac{\text{BR}(B^+ \to K^{ +} \pi^+ D_s^{(\ast)-})}
{\text{BR}(B^+ \to \pi^{+} \pi^+ D^{(\ast )-})} \sim 0.13 \ , 
\end{equation}
following the rule of analogy. Here the value of 0.13 for the latter ratio is taken as the experimental average~\cite{Tanabashi:2018oca}. Then we have BR$(B^0\to K^{\ast 0} D_s^{(\ast) +} D_s^{(\ast)-}) \sim 1.6\times 10^{-3}$.

The $b\to c \tau\nu$ background $B^0\to K^{\ast 0}D_s^{(\ast)-}\tau^+\nu$ is suppressed by both of the $s\bar{s}$ and $\tau$ productions. To estimate its rate, we introduce $B^+ \to K^+ D_s^{(\ast)-} \ell^+ \nu$ as a reference, with its branching ratio being known to be $6.1 \times 10^{-4}$~\cite{Stypula:2012mf}. This reference process suffers a suppression caused by the $s\bar{s}$ production also. The suppression factor for $B^0\to K^{\ast 0}D_s^{(\ast)-}\tau^+\nu$ thus can be approximately calculated based on the phase spaces available to these two decays respectively. Notably, the decay of $B^+ \to K^+ D_s^{(\ast)-} \ell^+ \nu$ is dominated by a hadronic resonance with its mass near the $m_{K^+}+ m_{D_s^{(\ast)-}}$ threshold~\cite{Stypula:2012mf}. We assume so for $B^0\to K^{\ast 0} D_s^{(\ast)-}\tau^+ \nu$ with the correspondent resonance mass being near the $ m_{K^{\ast 0}}+ m_{D_s^{(\ast)-}}$ threshold instead~\footnote{As a conservative treatment, we assume such decays to be three-body (the hadronic resonance with its mass right at the said threshold, the charged lepton and the neutrino) for evaluating the suppression caused by the $\tau$ production. Then we have the differential variable dPS$_3 \propto M^{-3} dm_{12}^2 dm_{23}^2$~\cite{Tanabashi:2018oca}. Here $M$ is the invariant mass of the parent particle and $m_{ij}$ is the invariant mass of the $i$-th and $j$-th particles among the three. With an  assumption of four-body decay, the suppression is typically evaluated to be one order stronger. }. As for $B_s\to \bar{K}^{\ast 0} D^{(\ast)-}\tau^+\nu$, it is not suppressed by the $s\bar{s}$ production and thus has a larger rate compared to $B^0\to K^{\ast 0}D_s^{(\ast)-}\tau^+\nu$. Its rate can be estimated similarly, using BR$(B_s \to D_{s1}(2536)^-\mu^+\nu, {\rm with} \ D_{s1}(2536)^-\to \bar{K}_S^0  D^{\ast -} )=2.6\times 10^{-3}$~\cite{Tanabashi:2018oca} as a reference.

\begin{table}[h]
\centering
\begin{tabular}{c|ccccc}
\hline
Type & Channel & Color & $s\bar{s}$ & $\tau$ & BR  \\
\hline 
\hline 
\multirow{3}{1.5cm}{$b\to c\bar{c}s$} &$B_s \to\phi  D_s^{(\ast)+} D_s^{(\ast)-} $ & & \checkmark & & $ 1.6\times 10^{-3}$    \\ 
&$B^0 \to \phi D_s^{(\ast)+} D^{(\ast)-}  $ & &\checkmark & &  $ 1.6\times 10^{-3}$   \\ 
&$B_s \to \phi D^{(\ast)+} D^{(\ast)-}  $ & \checkmark & & &  $3.2\times 10^{-3}$    \\ 
\hline 
\multirow{2}{1.5cm}{$b\to c\tau\nu$} &$B_s\to \phi D_s^{(\ast)-}\tau^+ \nu$ & & \checkmark & \checkmark &  $2.6 \times 10^{-5}$   \\ 
& $B^0 \to \phi D^{(\ast)-} \tau^+ \nu$ & & \checkmark & \checkmark &  $< 4.0\times 10^{-7}$    \\
\hline
\end{tabular} 
\caption{Major backgrounds for the $B_s\to\phi \tau^+ \tau^-$ measurement. ``Color'', ``$s\bar{s}$'' and ``$\tau$'' represent the three effects suppressing the $B$-meson decays which are discussed in the text.}
\label{tab:Bsphibackground}
\end{table}  

The major backgrounds for the $B^0\to \phi \tau^+ \tau^-$ measurement are summarized in Tab.~\ref{tab:Bsphibackground}. To estimate the rate of $B_s \to \phi D^{(\ast)+} D^{(\ast)-}$, we take BR$(B^0\to K^{\ast 0} D^{(\ast)0} \bar{D}^{(\ast)0})=3.8\times 10^{-3}$~\cite{Tanabashi:2018oca} as a reference and assume 
\begin{equation}
\frac{\text{BR}(B_s \to \phi D^{(\ast)+} D^{(\ast)-}  )}{\text{BR}(B^0\to K^{ 0} D^{(\ast)0} \bar{D}^{(\ast)0})} = \frac{\text{BR}(B_s\to \phi X_{c\bar{c}})}{\text{BR}(B^0\to K^{ 0} X_{c\bar{c}})} \sim 0.85~,
\label{eq:charmoniumratio}
\end{equation}
following the rule of analogy. Here all of these decays are color-suppressed. $X_{c\bar{c}}$ is a light charmonium, namely $J/\psi,~\eta_c$, $\chi_{c1}$ or $\psi(2S)$, and the value of 0.85 is taken as the average over these four $X_{c\bar{c}}$ possibilities~\cite{Amhis:2019ckw,Tanabashi:2018oca}. Then one can find BR$(B_s \to \phi D^{(\ast)+} D^{(\ast)-}  ) \sim 3.2 \times 10^{-3}$. The remaining two $b\to c\bar{c}s$ backgrounds, namely $B_s \to\phi  D_s^{(\ast)+} D_s^{(\ast)-}$ and $B^0 \to \phi D_s^{(\ast)+} D^{(\ast)-}$, are not color-suppressed, but both suppressed by the $s\bar{s}$ production. This property is also shared by the decay of  $B^0\to K^{\ast 0} D_s^{(\ast)+} D_s^{(\ast)-}$. Thus, we simply assume that these three modes have the same branching ratios.

The two $b\to c\tau\nu$ backgrounds are suppressed by both of the $s\bar{s}$ and $\tau$ productions. To estimate the rate of $B_s\to \phi D_s^{(\ast)-}\tau^+ \nu$, again we take $B^+ \to K^+ D_s^{(\ast)-} \ell^+ \nu$ as a reference. Then its suppression factor can be approximately calculated based on the phase spaces available to these two decays respectively.  The estimation of BR$(B^0 \to \phi D^{(\ast)-} \tau^+ \nu)$ is more involved, as one needs to estimate BR$(B^0 \to \phi D^{(\ast)-} \ell^+ \nu)$ first. For this purpose, we take BR$(B^0 \to  \pi^+\pi^- D^{(\ast)-} \ell^+ \nu)=2.7\times 10^{-3}$~\cite{Amhis:2019ckw} as a reference and assume 
\begin{equation}
\frac{\text{BR}(B^0 \to \phi D^{(\ast)-} \ell^+ \nu)}{\text{BR}(B^0 \to  \pi^+\pi^- D^{(\ast)-} \ell^+ \nu)} = \frac{\text{BR}(B^0\to\phi \bar{D}^0)}{\text{BR}(B^0 \to  \pi^+\pi^-\bar{D}^0)} < 0.0023~.
\end{equation}
Here the upper bound is taken from the experimental average~\cite{Amhis:2019ckw}. Then we find BR($B^0 \to \phi D^{(\ast)-} \tau^+ \nu) < 4.0\times 10^{-7}$, a bound orders stronger than the branching ratios of the other listed backgrounds. This outcome can be explained by the OZI rule~\cite{Geiger:1991ab}, that is, the $\phi$ meson in $B^0 \to \phi D^{(\ast)-} \tau^+ \nu$, $B^0 \to \phi D^{(\ast)-} \ell^+ \nu$ and $B^0\to\phi \bar{D}^0$ can not be generated at leading order by single gluon emission.

\begin{table}[h]
\centering
\begin{tabular}{c|ccccc}
\hline
Type & Channel & Color & $s\bar{s}$ & $\tau$ & BR  \\
\hline 
\hline 
\multirow{2}{1.5cm}{$b\to c\bar{c}s$} &$B^+ \to K^+ D^{(\ast)+} D^{(\ast)-}$ & \checkmark &  &  & $2.8\times 10^{-3}$~\cite{Tanabashi:2018oca}  \\ 
&$B^+ \to K^+ D_s^{(\ast)+} D_s^{(\ast)-} $ & & \checkmark &  &  $1.6\times 10^{-3}$\\ 
\hline 
\multirow{1}{1.5cm}{$b\to \tau \nu$} & $B^+ \to K^+ D_s^{(\ast)-}  \tau^+ \nu$ & & \checkmark & \checkmark &  $  9.5\times 10^{-5}$   \\ 
\hline
\end{tabular} 
\caption{Major backgrounds for the $B^+ \to K^+ \tau^+ \tau^- $ measurement. ``Color'', ``$s\bar{s}$'' and ``$\tau$'' represent the three effects suppressing the $B$-meson decays which are discussed in the text.}
\label{tab:BcKcbackground}
\end{table}

The major backgrounds for the $B^+ \to K^+ \tau^+ \tau^-$ measurement are summarized in Tab.~\ref{tab:BcKcbackground}. Among the two $b\to c\bar{c}s$ backgrounds, $B^+ \to K^+ D^{(\ast)+} D^{(\ast)-}$ has been experimentally measured~\cite{Tanabashi:2018oca}. $B^+ \to K^+ D_s^{(\ast)+} D_s^{(\ast)-} $ is not color-suppressed, but suppressed by the $s\bar{s}$ production. This property is also shared by $B_s \to\phi  D_s^{(\ast)+} D_s^{(\ast)-} $ and $B^0\to K^{\ast 0} D_s^{(\ast)+}D_s^{(\ast)-}$. Thus we assume that they have the same branching ratios. The $b\to c\tau\nu$ background, namely $B^+ \to K^+ D_s^{(\ast)-}  \tau^+ \nu$, is suppressed by both of the $s\bar{s}$ and $\tau$ productions. Its rate can be estimated in the same way as we did for $B_s\to \phi D_s^{(\ast)-}\tau^+ \nu$ above.

\begin{table}[h!]
\centering
\begin{tabular}{c|ccccc}
\hline
Type & Channel & Color & $s\bar{s}$ & $\tau$ & BR  \\
\hline
\hline 
\multirow{2}{1.5cm}{$b\to c\bar{c}s$} & $B_s \to D_s^{(\ast)+} D_s^{(\ast)-}$ & & & &  $3.2 \times 10^{-2}$~\cite{Amhis:2019ckw} \\ 
&$B^0 \to D_s^{(\ast)+} D^{(\ast)-}$ & & & &  $4.0\times 10^{-2}$~\cite{Tanabashi:2018oca} \\ 
\hline 
\multirow{2}{1.5cm}{$b\to c\tau \nu$}  &$B_s \to D_s^{(\ast)-} \tau^+\nu$ & & & \checkmark &  $2.7\times 10^{-2}$ \\ 
&$B^0 \to D^{(\ast)-} \tau^+\nu$ &  & & \checkmark &  $2.7\times 10^{-2}$~\cite{Tanabashi:2018oca} \\
\hline 
\multirow{4}{1.5cm}{$b\to c\bar{u}d$} &$B_s\to D_s^{(\ast)-} \pi^+\pi^+\pi^- $ & & & &  $3.9\times 10^{-3}$ \\  
&$B^0\to D^{(\ast)-}  \pi^+\pi^+\pi^- $ & & &  & $3.9\times 10^{-3}$~\cite{Tanabashi:2018oca} \\
&$B^0\to D^{(\ast)-} a_1^+$ & & & &  $1.9\times 10^{-2} $~\cite{Tanabashi:2018oca} \\
&$B_s\to D_s^{(\ast)-} a_1^+$ & & & &  $1.9\times 10^{-2}$\\
\hline
\end{tabular} 
\caption{Major backgrounds for the $B_s \to \tau^+ \tau^-$ measurement. ``Color'', ``$s\bar{s}$'' and ``$\tau$'' represent the three effects suppressing the $B$-meson decays which are discussed in the text. The $\pi^\pm\pi^\pm\pi^\mp$ appearing in the first two $b\to c\bar{u}d$ channels are produced non-resonantly. }
\label{tab:Ditaubackground}
\end{table}

The major backgrounds for the $B_s \to \tau^+ \tau^-$ measurement are summarized in Tab.~\ref{tab:Ditaubackground}. Aside from the $b \to c\bar{c}s $ and $b\to c\tau\nu$ backgrounds, the $b\to c\bar{u}d$ decays also become important. As discussed above, in this context the $B_s$ decay vertex is unknown and hence there is a good chance for the three charged pions, either from the $B$-meson decay directly or from the intermediate $a_1^\pm$ decay, to fake $\tau$ lepton also.  Different from the other three cases, the major backgrounds for this measurement have been mostly measured~\footnote{In the LHCb measurement~\cite{Aaij:2017xqt}, the total backgrounds are estimated using the data from control region.}. The three exceptions are of $B_s \to D_s^{(\ast)- }+ X$ type, with $X =  \tau^+\nu$, $a_1^+$ and $\pi^+\pi^+\pi^-$. Here we simply assume that they have the same branching ratios as their respective $B^0 \to D^{(\ast)- }+ X$ counterparts.

\subsection{More Kinematic Features}
\label{ssec:feature}

In addition to the invariant mass of the reconstructed $B$ mesons, we can use sub-$B$ kinematics, namely the kinematics of the $B$-meson decay products, to improve the separation between the signal and its backgrounds. Below we will consider three classes of such observables.

\begin{figure}[h!]
\centering
\includegraphics[height=6 cm]{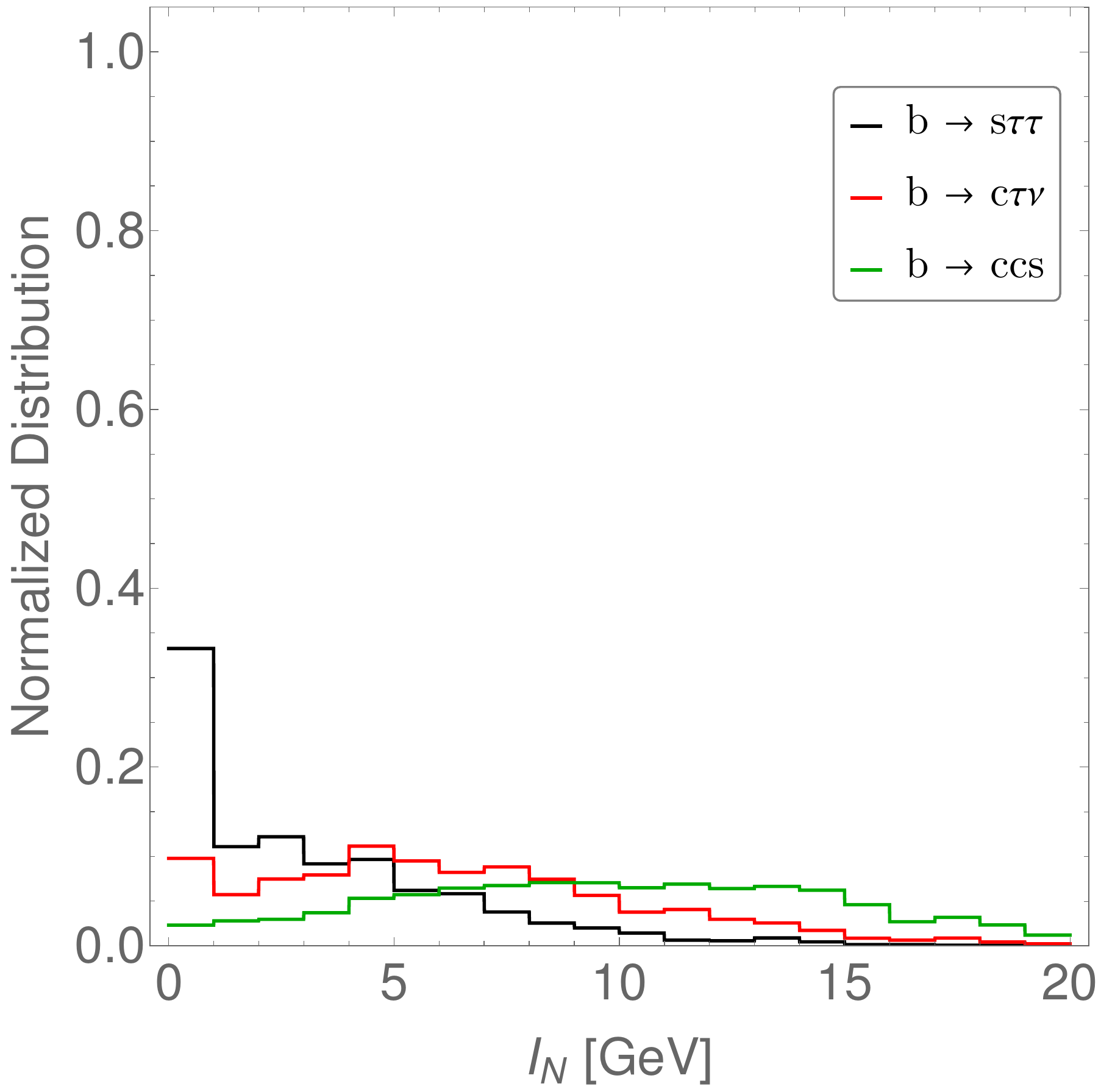}  \ \ \  \   \ \ 
\includegraphics[height=6 cm]{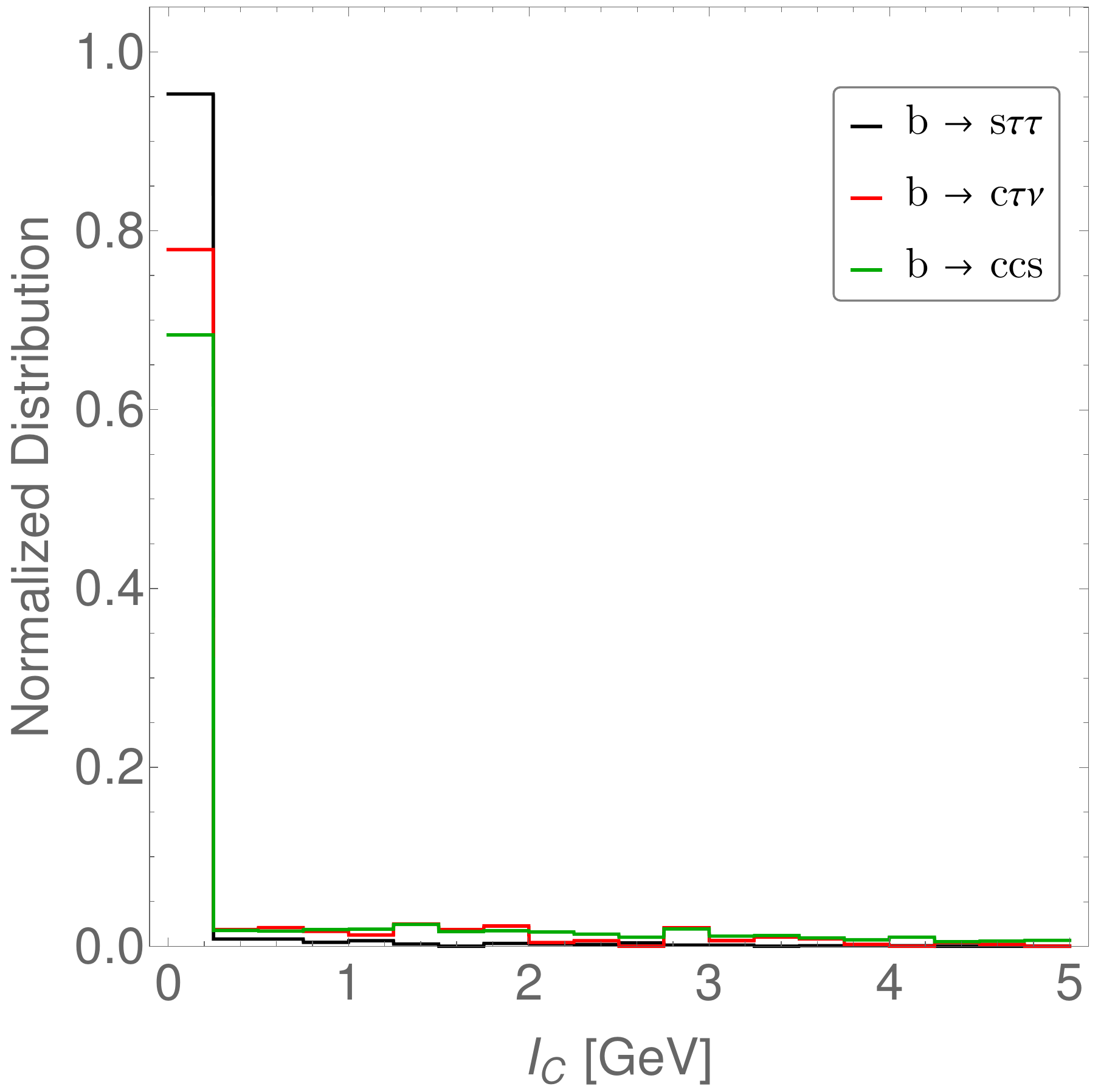}

\caption{Distributions of $I_N$ (left) and $I_C$ (right) for the $B^0\to K^{\ast 0} \tau^+ \tau^-$ measurement. Here $I_N$ and $I_C$ denote the overall isolation of the reconstructed $\tau$ leptons and $K^{\ast 0}$ meson in each event, from the other neural particles and displaced tracks, respectively.} 
\label{fig:isolation}
\end{figure}

The first class is the isolation of the intermediate particles reconstructed in each event, including $\tau$ leptons and $K^{\ast 0}(\phi,K^+)$ meson. As shown in Tab.~\ref{tab:ddecay}, the $D$-meson decays can produce neutral particles in addition to the charged mesons anticipated to mimic the $\tau$-lepton decay. Extra displaced tracks can be produced also, since $K_S^0$ meson as one of such neutral particles may decay to $\pi^+\pi^-$ with a relatively long lifetime. Isolating the reconstructed intermediate particles in each event thus may help suppress the relevant backgrounds. For this purpose, we introduce two overall isolation observables, namely $I_{N}$ and $I_{C}$, to quantify these effects. $I_N$ is defined as the total energy of the neutral particles found with $\Delta\Omega\leqslant 0.2$ w.r.t. any of the reconstructed $\tau$ leptons and $K^{\ast 0}(\phi,K^+)$ meson in each event, while $I_C$ as the total energy of all extra displaced tacks with their distance to the PV being $>0.1$mm. For illustration, we show the distributions of $I_N$ and $I_C$ for the $B^0\to K^{\ast 0} \tau^+ \tau^-$ measurement in Fig.~\ref{fig:isolation}. As expected, the signal events tend to have a smaller $I_N$ value compared to the backgrounds. This should be in part credited to the clean data environment of the $Z$ factories. It ensures no significant background contaminations for the signal to be introduced. $I_C$ is relatively weak. But it can still remove some backgrounds sitting on its distribution tail, with almost no signal loss.

\begin{figure}[h!]
\centering
\includegraphics[height=6cm]{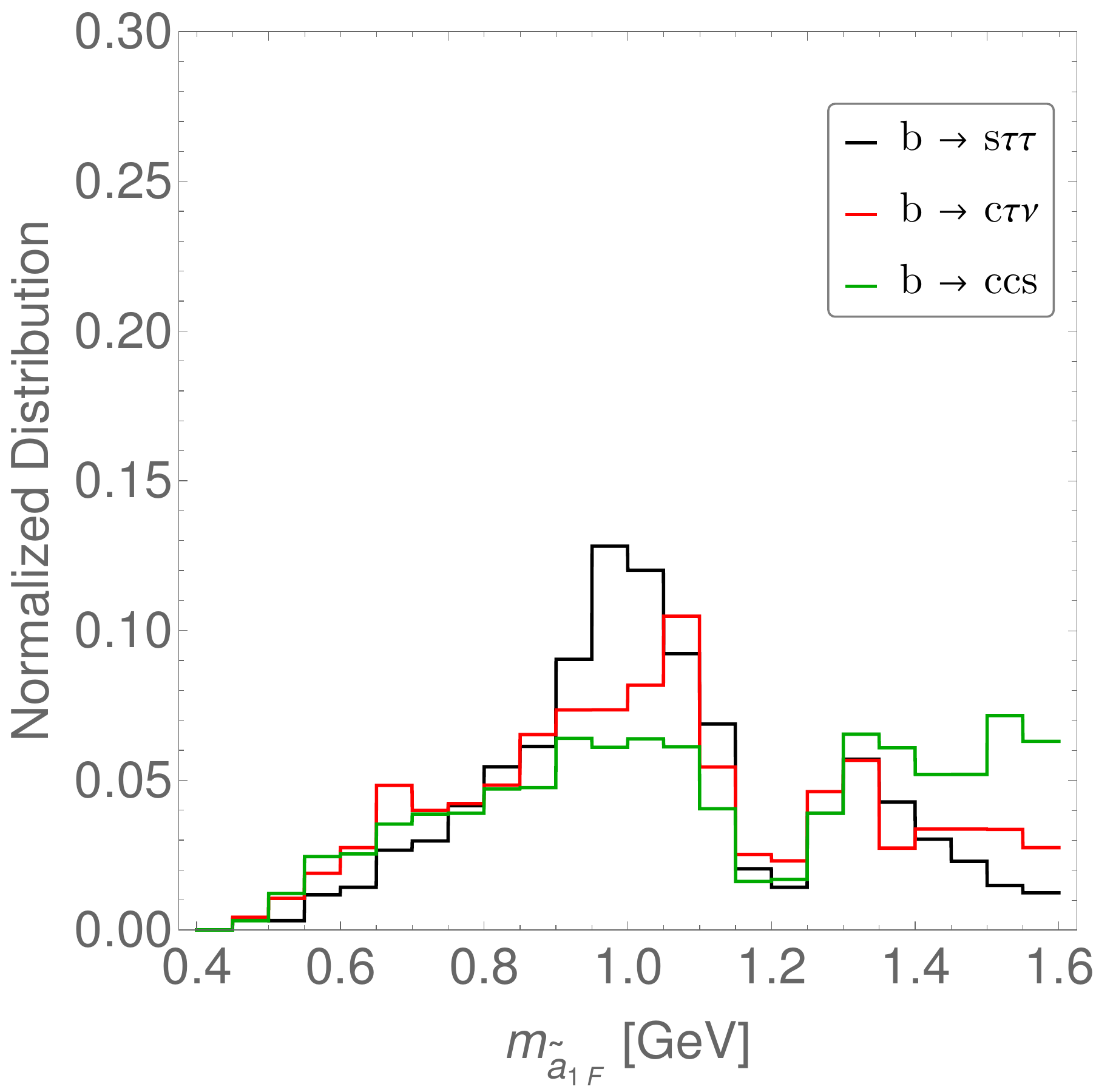}  \ \ \  \   \ \ 
\includegraphics[height=6cm]{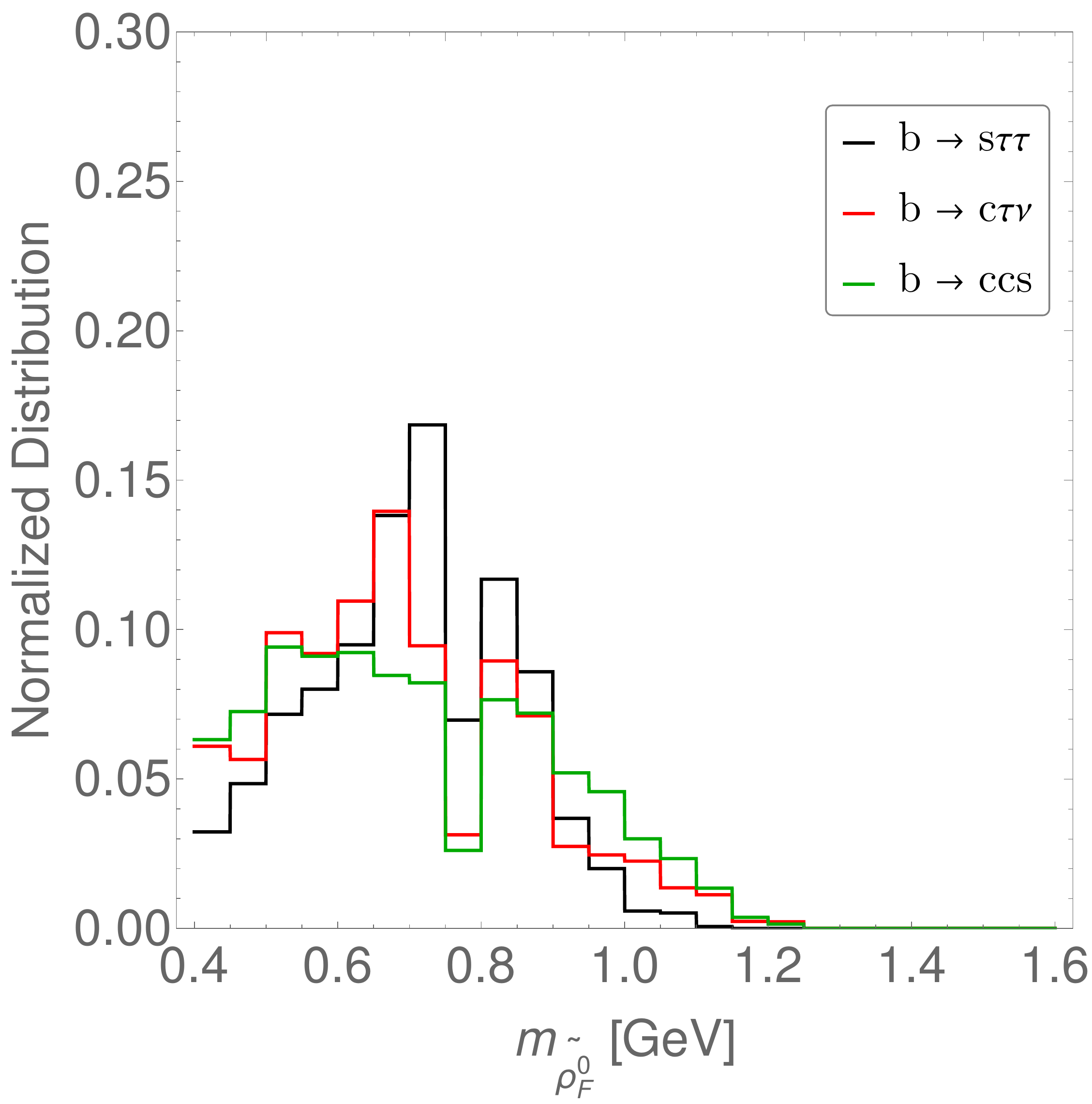}
\caption{Distributions of $m_{\tilde a_{1F}}$ (left) and $m_{\tilde \rho^0_F}$ (right) for the $B^0\to K^{\ast 0} \tau^+ \tau^-$ measurement. Here the subscript ``$F$'' represents one of the two reconstructed $a_1^\pm (\rho^0)$ in each event with its mass deviating more from the physical value.}
\label{fig:taurec1}
\end{figure}

The second class is the invariant mass of the intermediate particles reconstructed in each event, such as the $a_1^\pm$ resonance and $\rho^0$ meson appearing in the $\tau$-lepton decay chain. As discussed above, the decay mode of $\tau^\pm \to\pi^\pm\pi^\pm\pi^\mp\nu$ is realized mostly via the $a_1^\pm$ production and its subsequent decay to $\rho^0 \pi^\pm$~\cite{Asner:1999kj}. The invariant mass of the $\pi^\pm\pi^\pm\pi^\mp$ system is thus expected to be close to $m_{a_1}=1.2$GeV, and the invariant mass of at least one pion pair in this system to $m_{\rho^0} = 0.77$GeV. For illustration, we show the distributions of $m_{\tilde a_{1F}}$ and $m_{\tilde \rho^0_F}$ for the $B^0\to K^{\ast 0} \tau^+ \tau^-$ measurement in Fig.~\ref{fig:taurec1}. One can see that the background distributions are less convergent, compared to the signal ones.

\begin{figure}[h!]
\centering
\includegraphics[height=6cm]{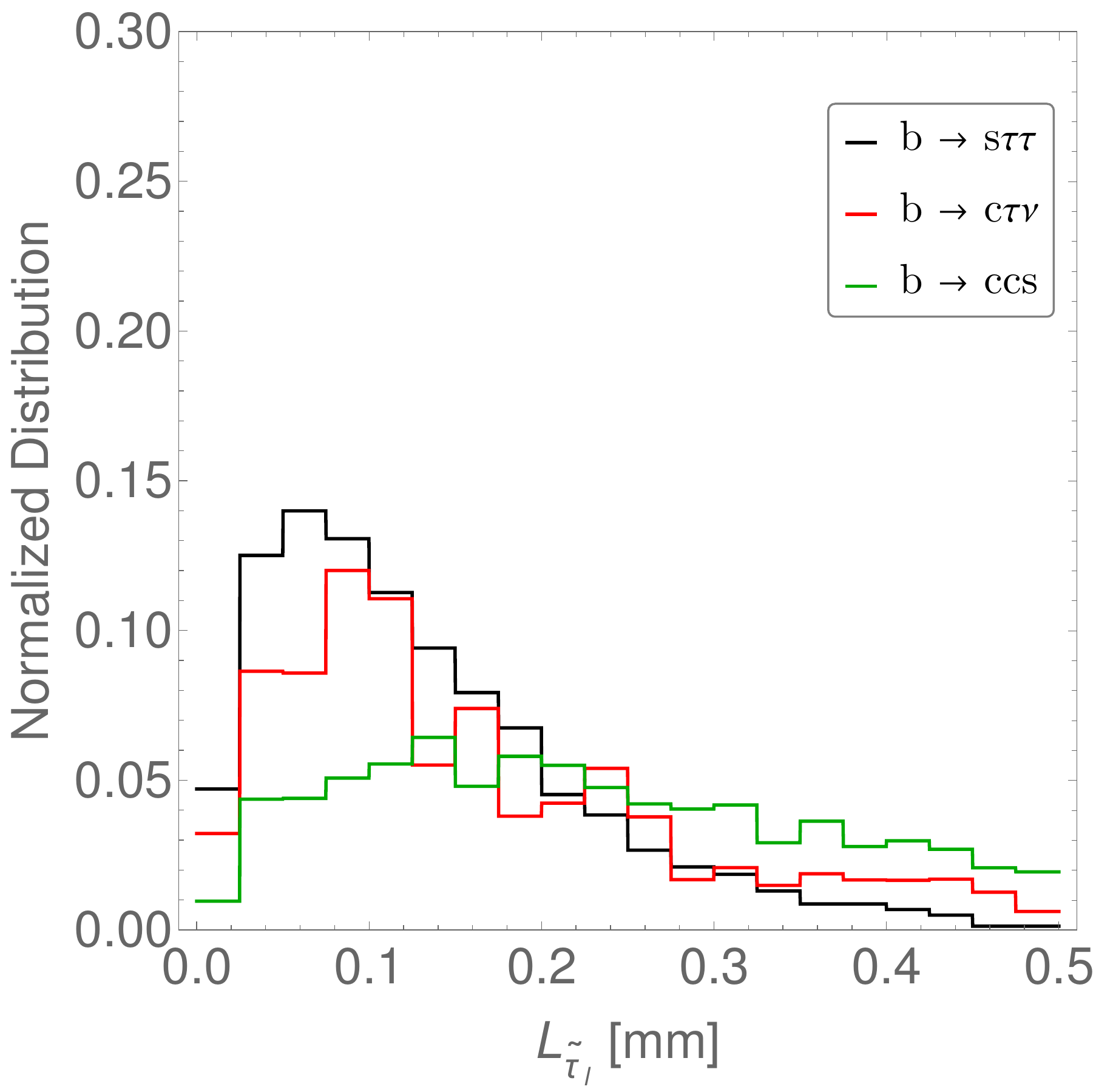}  \ \ \  \   \ \ 
\includegraphics[height=6cm]{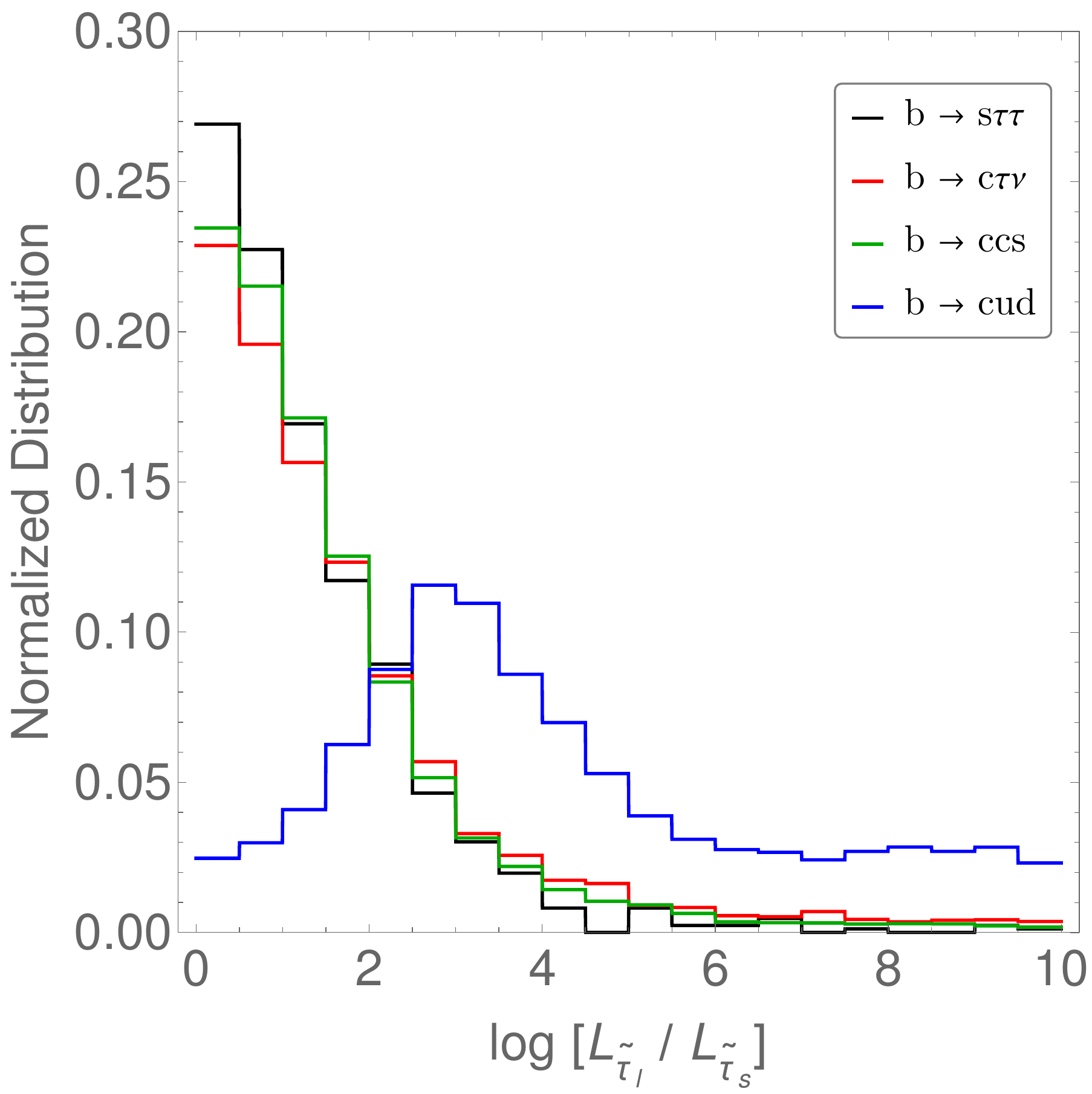}
\caption{Distributions of $L_{\tilde \tau}$ (left) for the $B^0\to K^{\ast 0} \tau^+ \tau^-$ measurement and $\log \frac{L_{\tilde \tau_l}}{L_{\tilde \tau_s}}$ (right) for the $B_s\to \tau^+ \tau^-$ measurement. Here $\tilde \tau_{l}$ and $\tilde \tau_{s}$ represent the reconstructed $\tau$ leptons in each event with the longer and shorter lifetimes, respectively.} 
\label{fig:taurec3}
\end{figure}

The third class is the proper decay length of the reconstructed $\tau$ leptons. As shown in the left panel of Fig.~\ref{fig:taurec3}, the backgrounds for the $B^0\to K^{\ast 0} \tau^+ \tau^-$  measurement tend to have a larger $L_{\tilde \tau} = |\vec{V}_{\tilde \tau}|/\gamma_{\tilde \tau}$, compared to the signal. Here $\gamma_{\tilde \tau}$ is the boost factor of $\tilde \tau$. This is because $D$ mesons, the potential $\tau$-lepton fakers, have a longer mean lifetime. For the $B_s\to \tau^+ \tau^-$ measurement, we can introduce another extra observable based on the $\tau$ proper decay length, namely $\log \frac{L_{\tilde \tau_l}}{L_{\tilde \tau_s}}$, to suppress its $b\to c\bar{u}d$  backgrounds. For such backgrounds, one of the two faked $\tau$ lepton in each event mostly arises from either the non-resonant  $\pi^\pm\pi^\pm\pi^\mp$ system or the $a_1^\pm$ resonance (see Tab.~\ref{tab:Ditaubackground}). It tends to overlap with the reconstructed $B_s$ vertex in space, and hence yield a $\tilde \tau_s$ value much smaller than the $\tilde \tau_l$ one. The feature is clearly displayed in the right panel of Fig.~\ref{fig:taurec3}.

\section{Analysis Results}
\label{sec:result}

Below we will analyze the sensitivities of measuring $B^0 \to K^{\ast 0}\tau^+\tau^-$, $B_s\to \phi \tau^+\tau^-$, $B^+ \to K^+ \tau^+ \tau^-$ and $B_s \to \tau^+\tau^-$ at the future $Z$ factories, using a cut-based method. Then the expected precisions will be applied to constrain the $b\to s\tau^+\tau^-$ operators in the EFT.

\subsection{Sensitivities at the Future $Z$ Factories}
\label{ssec:limit}

The samples for simulating these four $b\to s \tau^+\tau^-$ measurements are preselected via the steps detailed in Sec.~\ref{sec:pheno}. After that, three cuts based on the kinematics discussed in Subsec.~\ref{ssec:feature} are applied which include
\begin{itemize}

\item isolation cut (cut 1):  $I_N<0.5~\text{GeV}$ and $I_C < 4~\text{GeV}$.

\item  $a_1$ and $\rho$ mass-window cut (cut 2):  $ m_{\tilde{a}_{1F}} <1.5~\text{GeV}$ and $0.6<m_{\tilde{\rho}^0_F}<0.9 ~\text{GeV}$.

\item $\tau$ decay-length cut (cut 3): $L_{\tilde{\tau}_l}<0.25~\text{mm}$, and additionally $\log\frac{L_{\tilde{\tau}_l}}{L_{\tilde{\tau}_s}}<2$ for $B_s \to \tau^+\tau^-$ measurement.

\end{itemize}
At last, a mass window cut is applied for the reconstructed $B$ mesons, namely $|m_{\tilde B} - m_B| < 0.02$GeV for the first three $b\to s \tau^+\tau^-$ measurements and $m_{\tilde B_s} \in [5.0,6.0]$GeV for $B_s \to \tau^+\tau^-$.

\begin{table}[h!]
\centering
 \resizebox{\textwidth}{!}{  
\begin{tabular}{c|ccccccc} 
\hline
Process    & $N_{\rm evt}$   & $\epsilon_{\rm pre}$  & $\epsilon_{1}$  & $\epsilon_{2}$  & $\epsilon_{3}$  & $\epsilon_{m_{\tilde B}}$  & Tera-$Z$ Yield       \\ 
\hline
\hline 
$B\to K^{\ast 0}\tau^+\tau^-$  & $1.2\times 10^{4}$  & $3.3\times 10^{-3}$  & $6.3\times 10^{-1}$ & $7.3\times 10^{-1}$ & $8.8\times 10^{-1}$   & $6.4\times 10^{-1}$ & $1.0\times 10^{1}$     \\ 
\hline 
$b\to c\tau\nu$ & $1.8\times 10^{7}$  & $2.5\times 10^{-4}$ & $1.8\times 10^{-1}$  & $4.9\times 10^{-1}$ & $7.0\times 10^{-1}$ & $7.4\times 10^{-2}$ & $2.1\times 10^1$ \\
$b\to c\bar{c}s$  & $2.4\times 10^{9}$ & $2.9\times 10^{-4}$ & $4.2\times 10^{-2}$  & $4.0\times 10^{-1}$ & $4.7\times 10^{-1}$ & $4.5\times 10^{-2}$ & $2.4\times 10^2$ \\ 
\hline
\hline 
$B_s\to\phi \tau^+ \tau^-$ & $2.8\times 10^{3}$ & $2.0\times 10^{-3}$ & $6.3\times 10^{-1}$ & $7.3\times 10^{-1}$ & $8.8\times 10^{-1}$ & $6.3\times 10^{-1}$ & $1.4$ \\ 
\hline
$b\to c\tau\nu$ & $8.8\times 10^{5}$  & $4.2\times 10^{-4}$ & $2.4\times 10^{-1}$  & $5.4\times 10^{-1}$ & $8.1\times 10^{-1}$ & $1.2\times 10^{-1}$ & $4.7$ \\ 
$b\to c\bar{c}s$  & $3.5\times 10^{8}$ & $3.0\times 10^{-4}$ & $6.3\times 10^{-2}$  & $3.5\times 10^{-1}$ & $5.1\times 10^{-1}$ & $5.2\times 10^{-2}$ & $6.0\times 10^1$ \\ 
\hline 
\hline
$B^+ \to K^+ \tau^+ \tau^-$      & $1.4\times 10^{4}$  & $5.9\times 10^{-3}$  & $6.3\times 10^{-1}$ & $6.9\times 10^{-1}$ & $9.0\times 10^{-1}$   & $6.0\times 10^{-1}$ & $2.0\times 10^{1}$     \\ 
\hline
$b\to c\tau\nu$ & $1.1 \times 10^{7}$  & $3.3\times 10^{-3}$  & $2.1\times 10^{-1}$ & $4.7\times 10^{-1}$ & $7.9\times 10^{-1}$   & $2.6\times 10^{-2}$ & $7.5\times 10^{1}$    \\
$b\to c\bar{c}s$  & $5.3 \times 10^8$  & $2.3\times 10^{-3}$  & $6.2\times 10^{-2}$ & $3.2\times 10^{-1}$ & $6.9\times 10^{-1}$   & $2.8\times 10^{-2}$ & $4.5\times 10^{2}$    \\
\hline 
\hline 
$B_s \to \tau^+\tau^-$ & $2.5\times 10^{4}$   & $9.2\times 10^{-3}$  & $5.4\times 10^{-1}$ & $6.4\times 10^{-1}$ & $6.7\times 10^{-1}$   & $6.6\times 10^{-1}$ & $3.4\times 10^{1}$  \\  
\hline
$b\to c\tau\nu$ & $4.1\times 10^{9}$          & $5.4\times 10^{-3}$  & $2.1\times 10^{-1}$ & $5.0\times 10^{-1}$ & $5.0\times 10^{-1}$   & $1.6\times 10^{-1}$ & $1.9 \times 10^{5}$  \\  
$b\to c\bar{c}s$ & $5.8\times 10^{9}$         & $6.2\times 10^{-3}$  & $5.3\times 10^{-2}$ & $3.8\times 10^{-1}$ & $4.0\times 10^{-1}$   & $4.0\times 10^{-1}$ & $1.2\times 10^{5}$  \\  
$b\to c\bar{u}d$ & $3.5\times 10^{9}$         & $9.9\times 10^{-3}$  & $1.9\times 10^{-1}$ & $4.7\times 10^{-1}$ & $1.1\times 10^{-1}$   & $4.8\times 10^{-1}$ & $1.6\times 10^{5}$  \\  
\hline 
\end{tabular} 
}
\caption{Cut flows and Tera-$Z$ yields for the measurements of $B^0\to K^{\ast 0}\tau^+\tau^-$, $B_s\to \phi \tau^+\tau^-$, $B^+ \to K^+ \tau^+ \tau^-$ and $B_s \to \tau^+\tau^-$. $\epsilon$ denotes the efficiency for each cut defined in the text.}

\label{tab:resultks}
\end{table}

The cut flows and Tera-$Z$ yields for measuring $B^0\to K^{\ast 0}\tau^+\tau^-$, $B_s\to \phi \tau^+\tau^-$, $B^+ \to K^+ \tau^+ \tau^-$ and $B_s \to \tau^+\tau^-$ are summarized in Tab.~\ref{tab:resultks} (a more detailed version can be found in Appendix~\ref{app:cutflow}). From this table, one can see that the isolation (mainly $I_N$) cut is extremely efficient in removing the $b\to c\bar{c}s$ backgrounds. This is because the decays of $D$ mesons, which are contained more in the $b\to c\bar{c}s$ backgrounds, tend to produce more extra particles other than the $\pi^\pm \pi^\pm \pi^\mp$ used for $\tau$ reconstruction. The $a_1$ and $\rho$ mass-window cut is relatively weak, but it benefits the quality of the reconstructed $\tau$ leptons. The $\tau$ decay-length cut vetoes the events with the lifetime being incompatible with the real one for their reconstructed $\tau$ leptons. As expected from the right panel of Fig.~\ref{fig:taurec3}, it can efficiently remove the $b\to c \bar ud $ backgrounds for the $B_s \to \tau^+\tau^-$ measurement.

 \begin{figure}[h!]
\centering
\includegraphics[height=4.5cm]{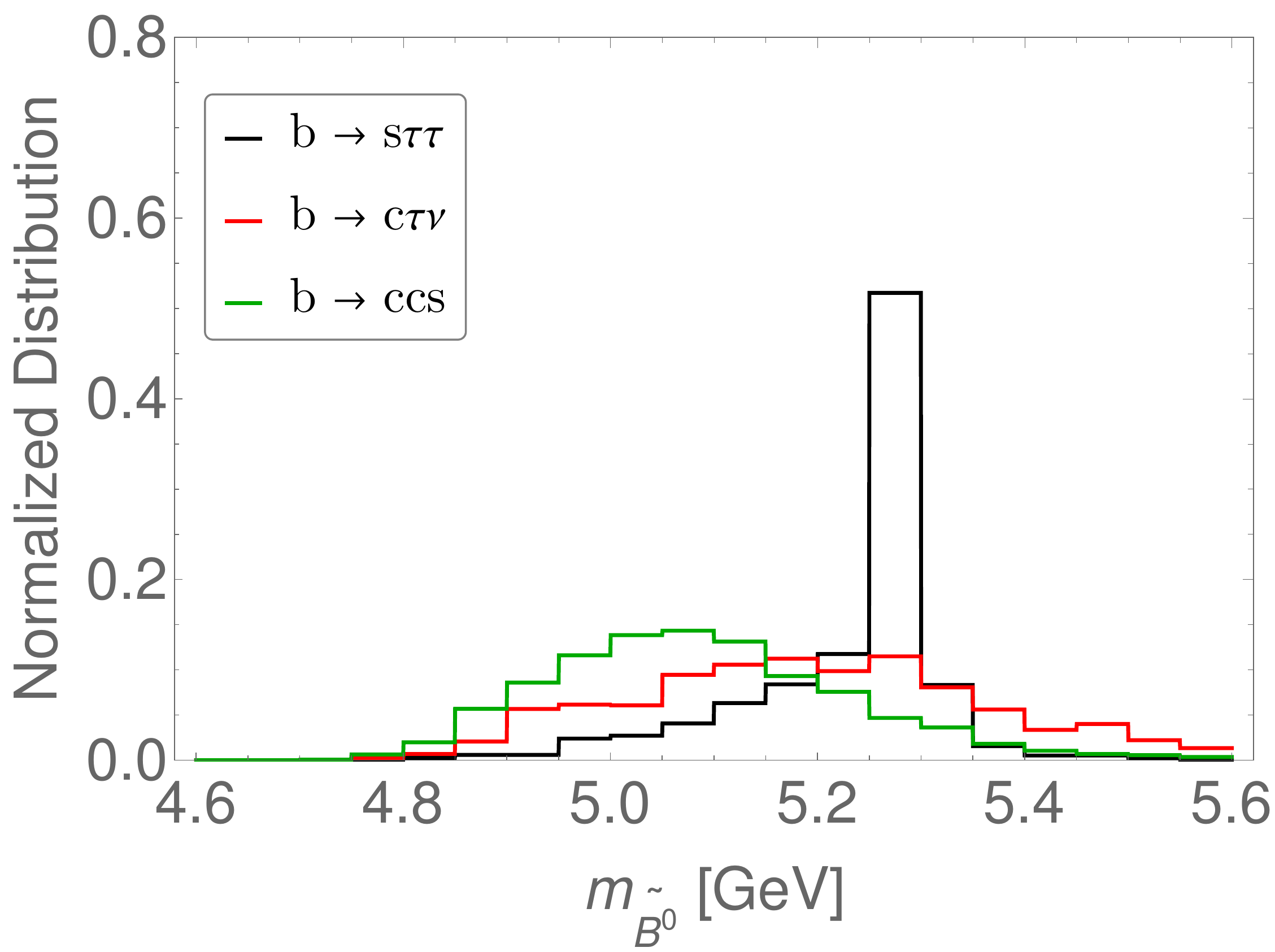} \ \  \ \  \ \ 
\includegraphics[height=4.5cm]{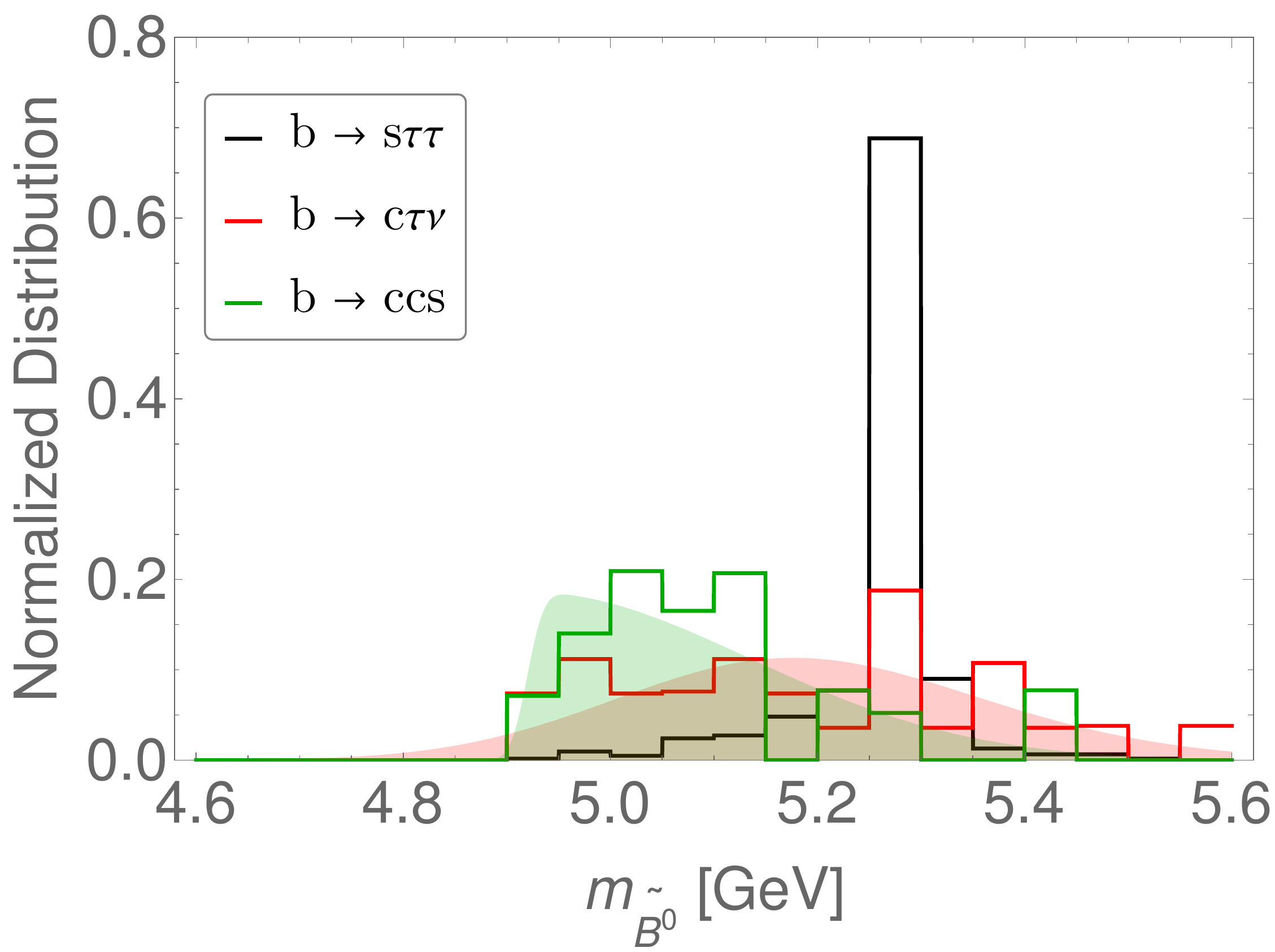}
\includegraphics[height=4.5cm]{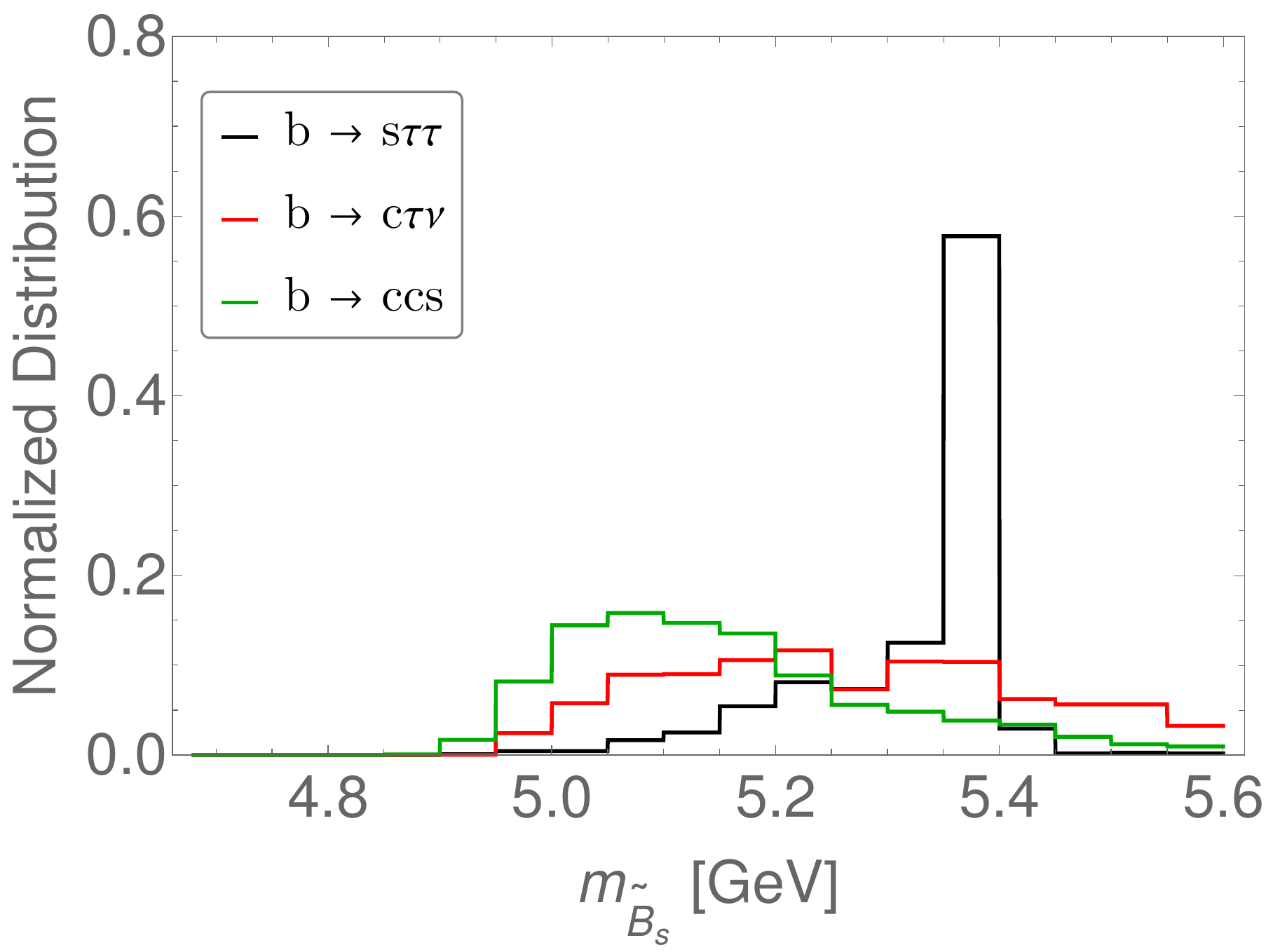} \ \ \ \ \ \   
\includegraphics[height=4.5cm]{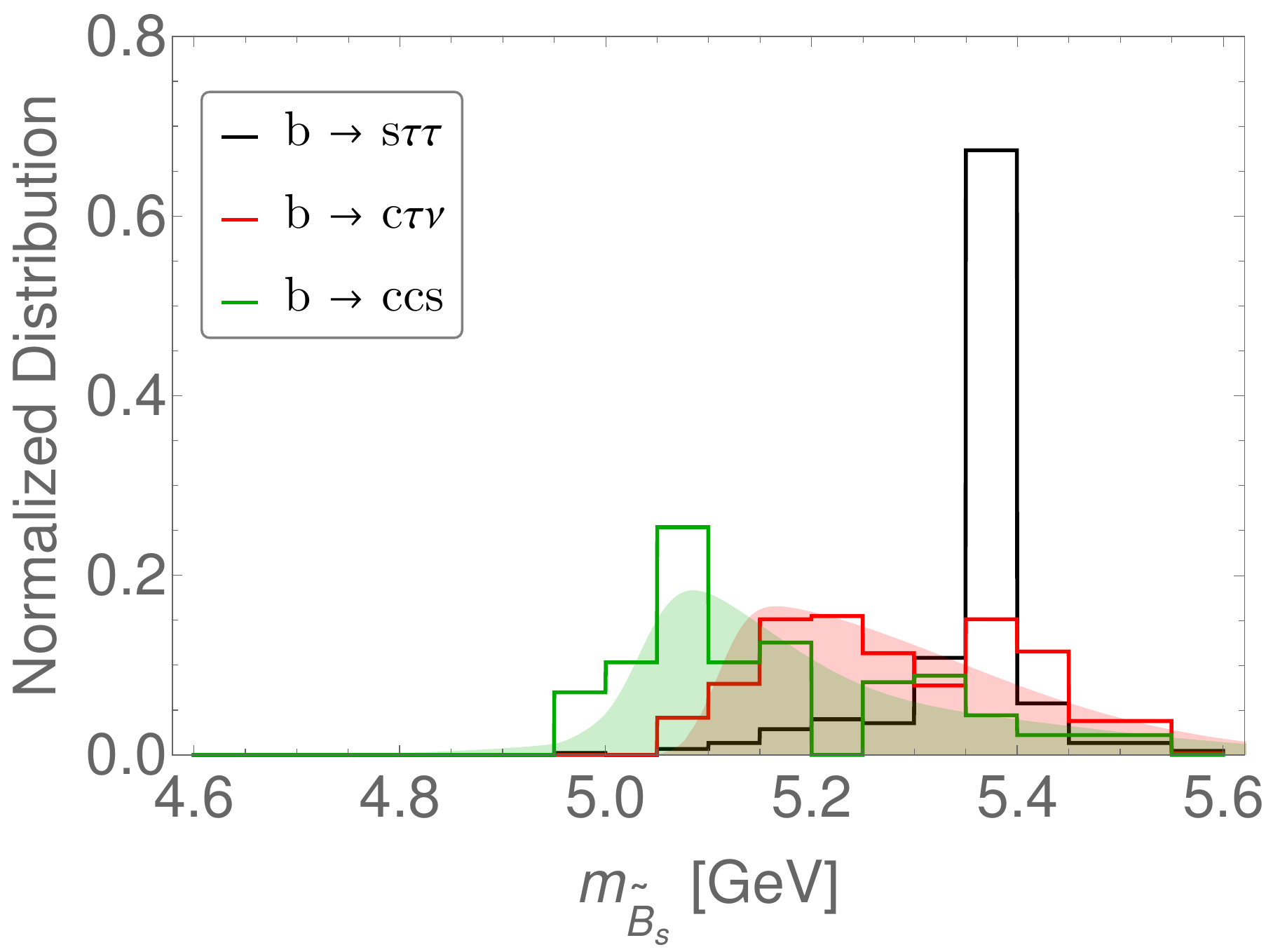}
\includegraphics[height=4.5cm]{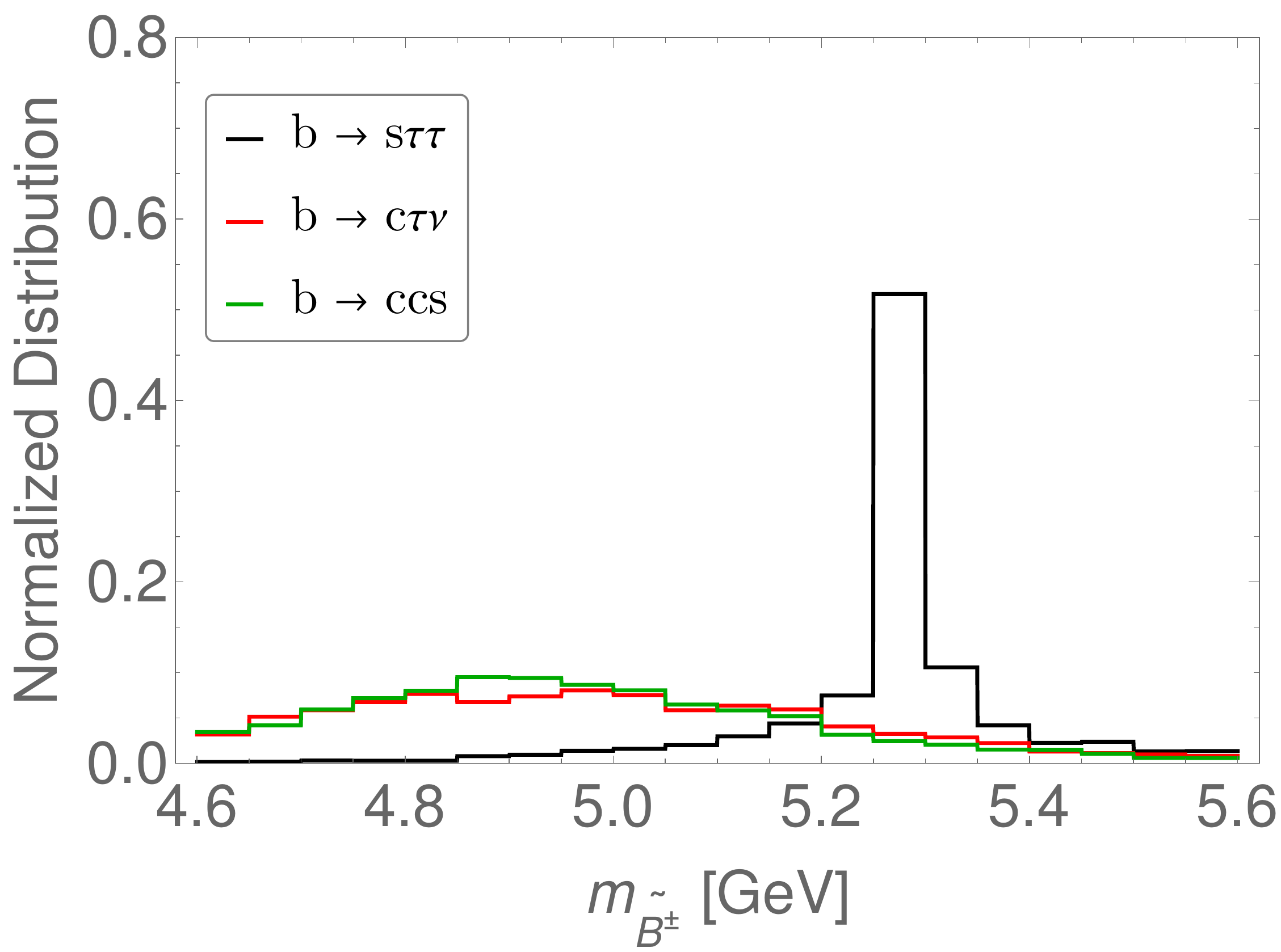} \ \ \ \ \ \   
\includegraphics[height=4.5cm]{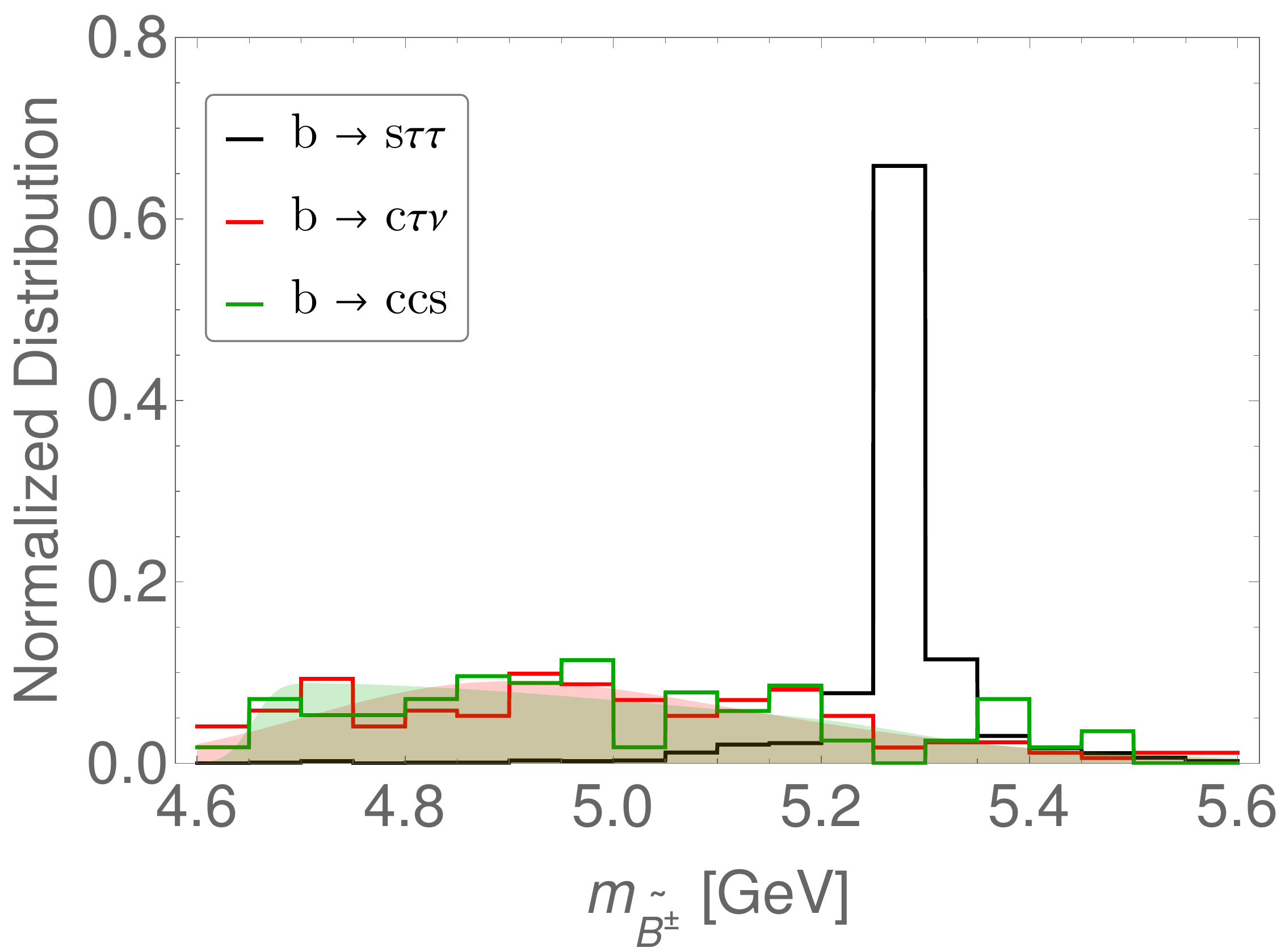}
\includegraphics[height=4.5cm]{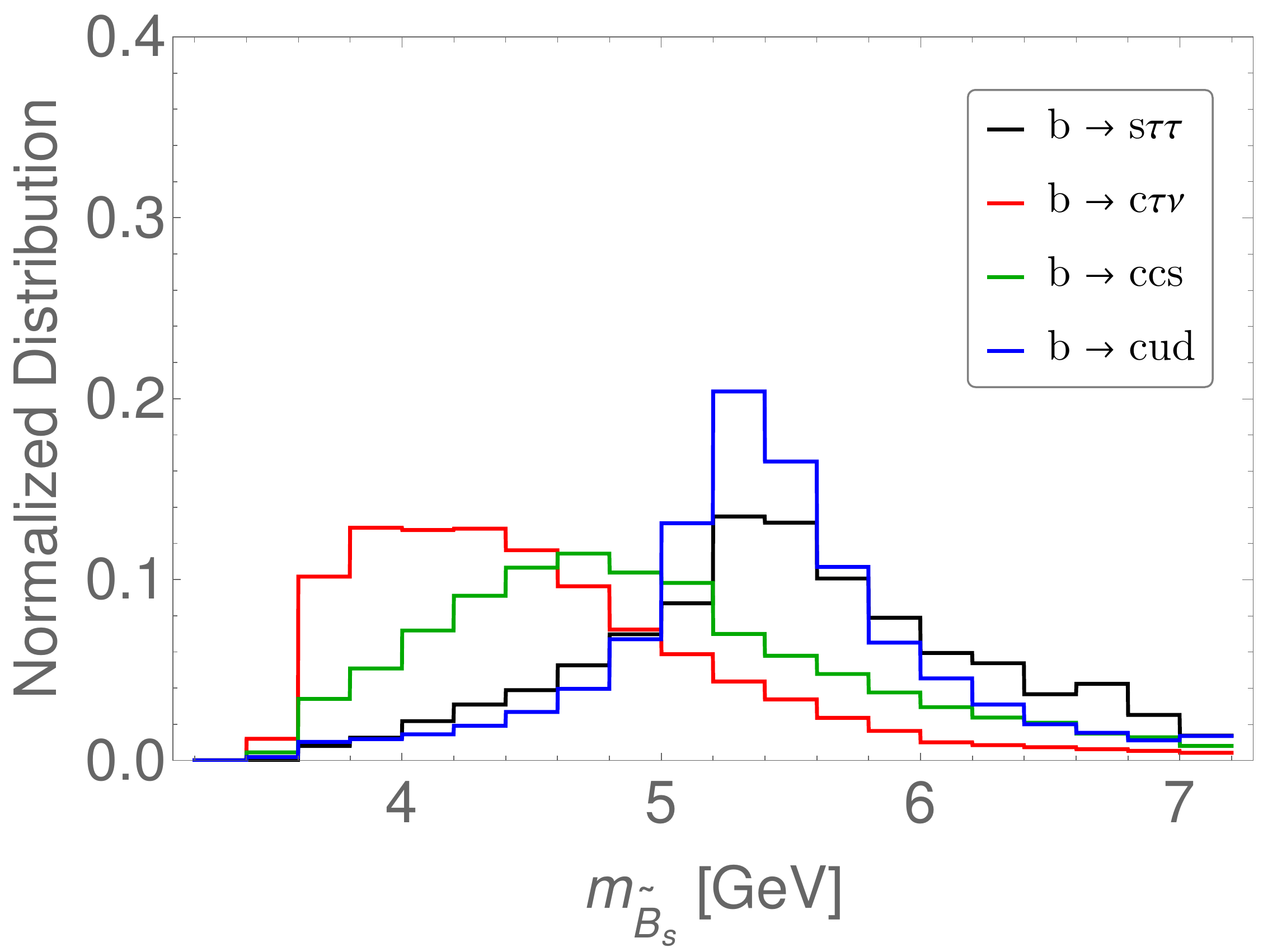} \ \  \ \  \ \ 
\includegraphics[height=4.5cm]{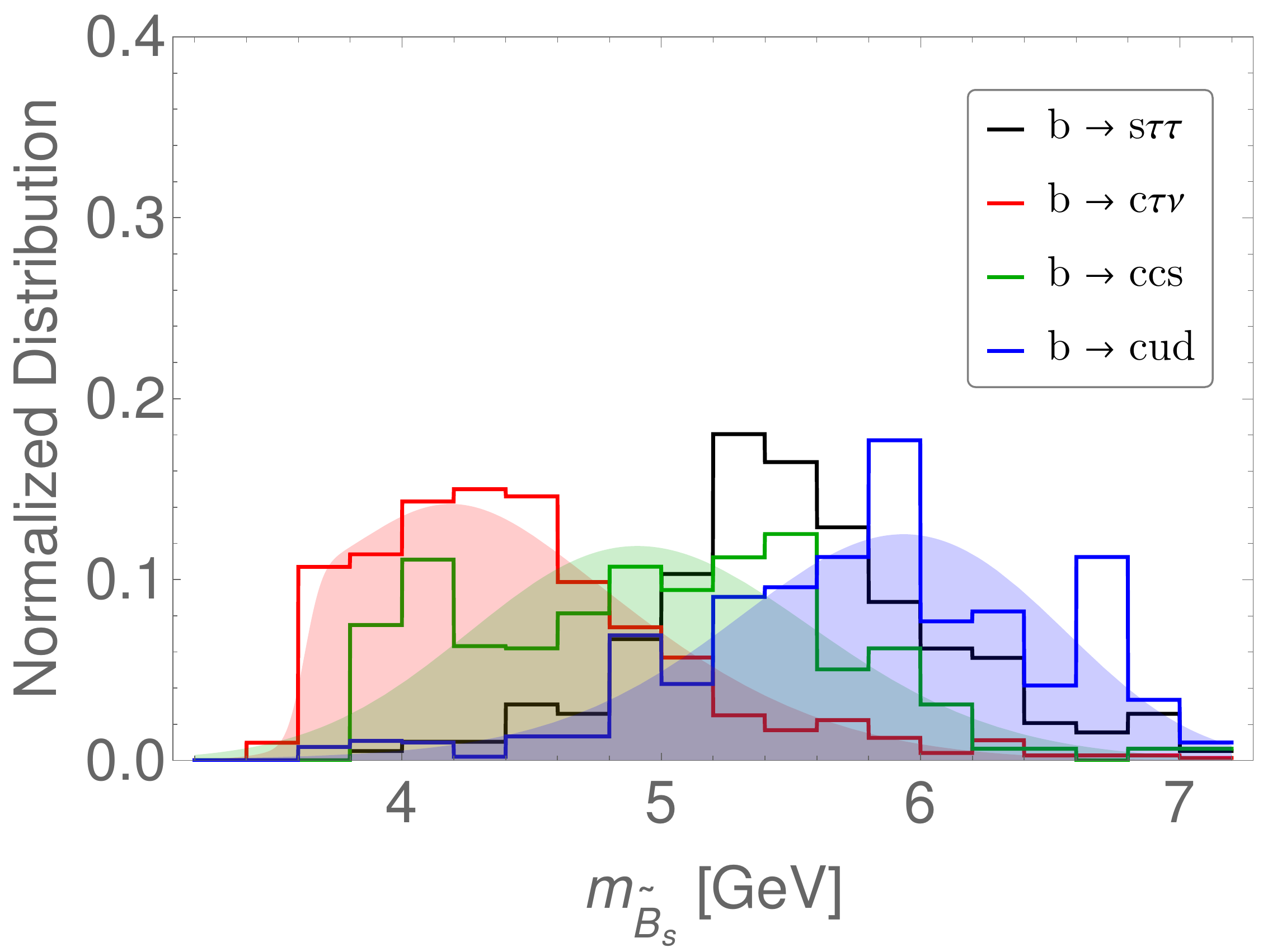}
\caption{Normalized distributions of the reconstructed $B$-meson mass for the measurements of $B^0\to K^{\ast 0} \tau^+ \tau^-$ (1st row) and $B_s\to\phi \tau^+ \tau^-$ (2nd row) and $B^+ \to K^+ \tau^+ \tau^-  $ (3rd row) and $B_s \to \tau^+ \tau^-$ (4th row), before (left) and after (right) cuts 1-3 are applied. In the right column, the shaded regions represent the background fittings using skew normal distribution.}
\label{fig:mtB}
\end{figure}

Fig.~\ref{fig:mtB} displays the normalized distributions of $m_{\tilde B}$ for these four measurements. The plots in the left and right columns are made before and after cuts 1-3, respectively. By comparing the two panels in each row, one can see that the ratio of signal and background, namely $S/B$, in the highest signal bin is improved to various extents by the applied cuts. But, due to their high rejections, the distributions of the backgrounds after these cuts suffer statistical fluctuations. To reduce the impacts of this effect on the analysis, we introduce skew normal distribution to fit them before applying the $m_{\tilde B}$ cut. Then $\epsilon_{\tilde B}$ is determined based on the skew normal distributions of these backgrounds.

\begin{table}[h!]
\centering
 \resizebox{\textwidth}{!}{  
\begin{tabular}{cccccc}
\hline 
Channel & Belle II~\cite{Kou:2018nap} & LHCb~\cite{Bediaga:2018lhg}  & Giga-$Z$ & Tera-$Z$ & $10 \times$Tera-$Z$   \\ 
\hline 
$B^0\to K^{\ast 0} \tau^+ \tau^-$ & - & - & $5.0^{+2.1}_{-1.5}(22,27)\times 10^{-6} $ & $1.6^{+0.7}_{-0.5}(6.8,8.5)\times 10^{-7}$  & $5.0^{+2.1}_{-1.5}(22,27)\times 10^{-8}$  \\ 
$B_s\to\phi \tau^+ \tau^-$ & - & - & $1.5^{+0.6}_{-0.4}(4.9,5.9) \times 10^{-5} $ & $4.8^{+1.9}_{-1.4}(15,19)\times 10^{-7}$ & $1.5^{+0.6}_{-0.4}(4.9,5.9)\times 10^{-7}$  \\ 
$B^+ \to K^+ \tau^+ \tau^-  $ & $<2.0\times 10^{-5}$ & - & $4.4^{+1.6}_{-1.1}(19,25)\times 10^{-6} $ & $1.4^{+0.6}_{-0.3}(6.0,8.0)\times 10^{-7}$ & $4.4^{+1.6}_{-1.1}(19,25)\times 10^{-8}$  \\ 
$B_s \to \tau^+ \tau^-$ &  $<8.1\times 10^{-4}$ & 5$\times 10^{-4}$ & $4.9^{+0.9}_{-0.7}(5.6,6.3)\times 10^{-4}$    & $1.5^{+0.3}_{-0.2}(1.8,2.0)\times 10^{-5}$  & $4.9^{+0.9}_{-0.7}(5.6,6.3)\times 10^{-6}$ \\ 
\hline 
\end{tabular}
}
\caption{Expected precisions ($@ 1\sigma$ C.L.) for the measurements of $B^0\to K^{\ast 0}\tau^+\tau^-$, $B_s\to \phi \tau^+\tau^-$, $B^+ \to K^+ \tau^+ \tau^-$ and $B_s \to \tau^+\tau^-$ at Belle~II, LHCb and the future $Z$ factories. The superscripts and subscripts for the numbers in the last three columns represent the precisions obtained by varying the experimentally measured backgrounds by one sigma and the semi-quantitatively estimated ones by a factor of two, upward and downward respectively. The two numbers in each parenthesis denote the sensitivities with a finite spatial resolution, $i.e.$, 5$\mu$m and 10$\mu$m respectively, for the tracker.}
\label{tab:comparison}
\end{table}

\begin{figure}[h!]
\centering
\includegraphics[height=7.5cm]{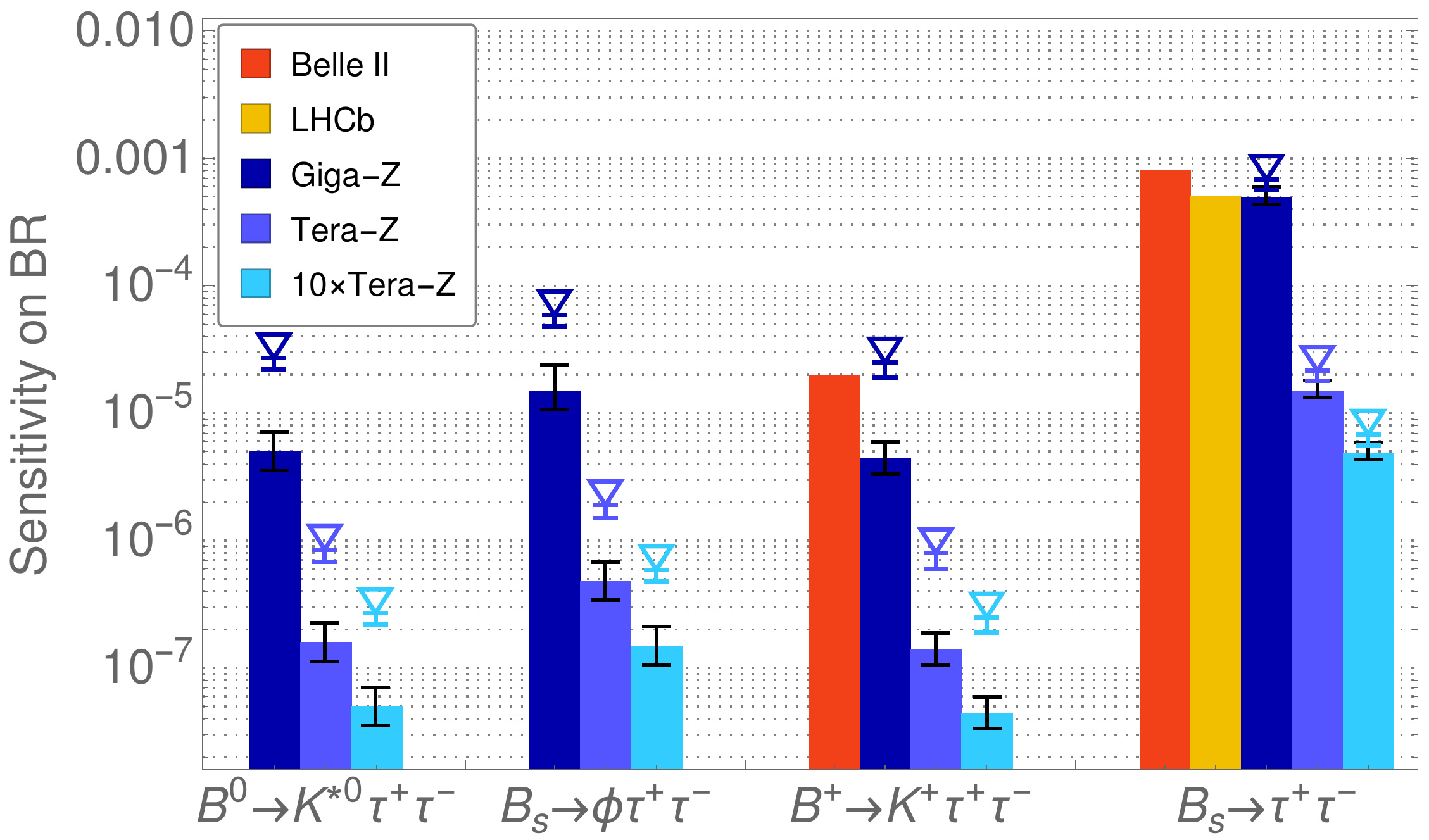}
\caption{Expected precisions ($@ 1\sigma$ C.L.) for the measurements of $B^0\to K^{\ast 0}\tau^+\tau^-$, $B_s\to \phi \tau^+\tau^-$, $B^+ \to K^+ \tau^+ \tau^-$ and $B_s \to \tau^+\tau^-$ at Belle~II, LHCb and the future $Z$ factories. The error bars represent the precisions obtained by varying the experimentally measured backgrounds by one sigma and the semi-quantitatively estimated ones by a factor of two, upward and downward respectively. The double bars below the inverted triangle denote the sensitivities with a finite spatial resolution, $i.e.$, 5$\mu$m and 10$\mu$m respectively, for the tracker.}
\label{fig:res}
\end{figure}

\begin{figure}[h!]
\centering
\includegraphics[height=6cm]{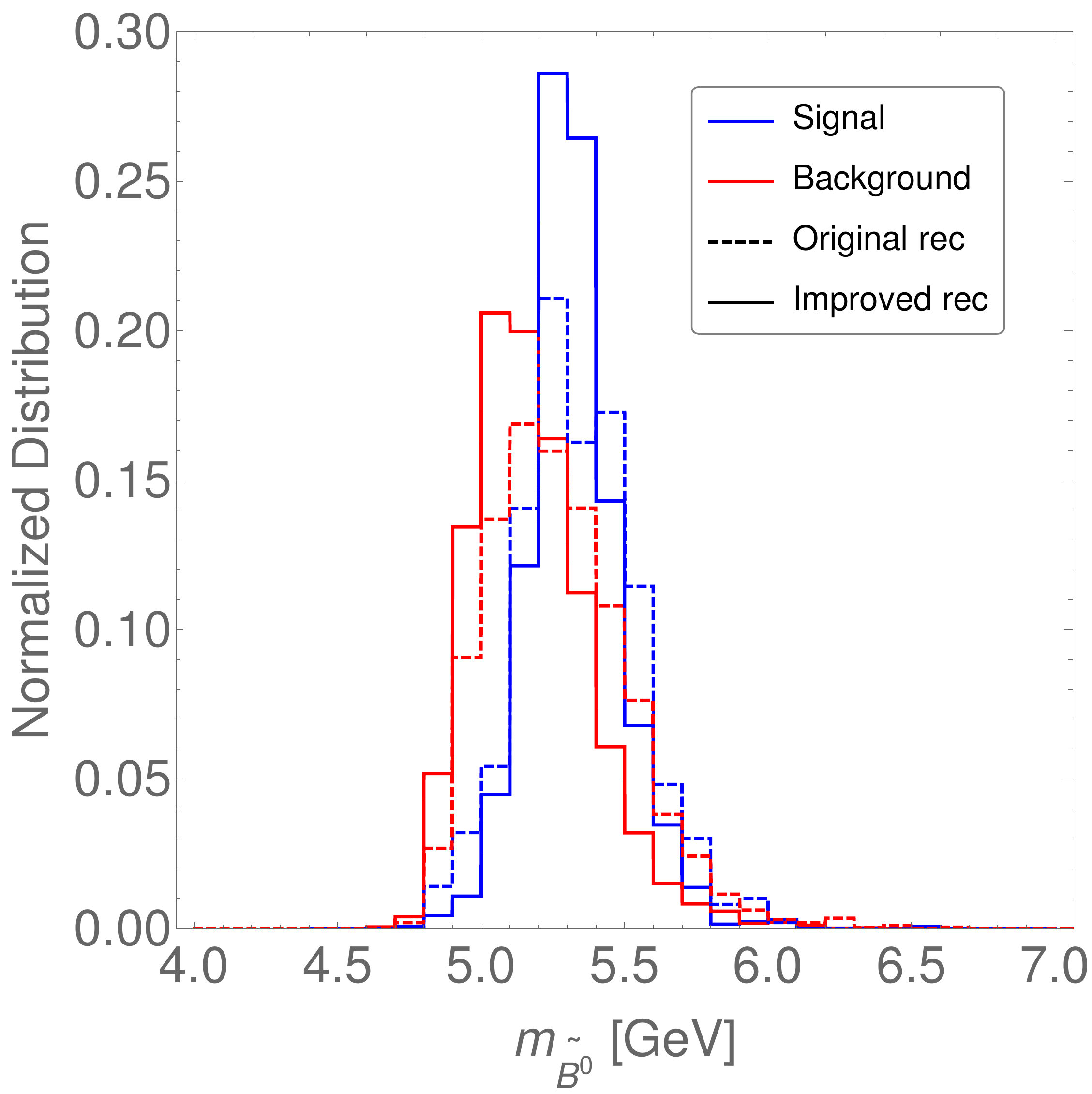} \ \ \  \   \ \ 
\includegraphics[height=6cm]{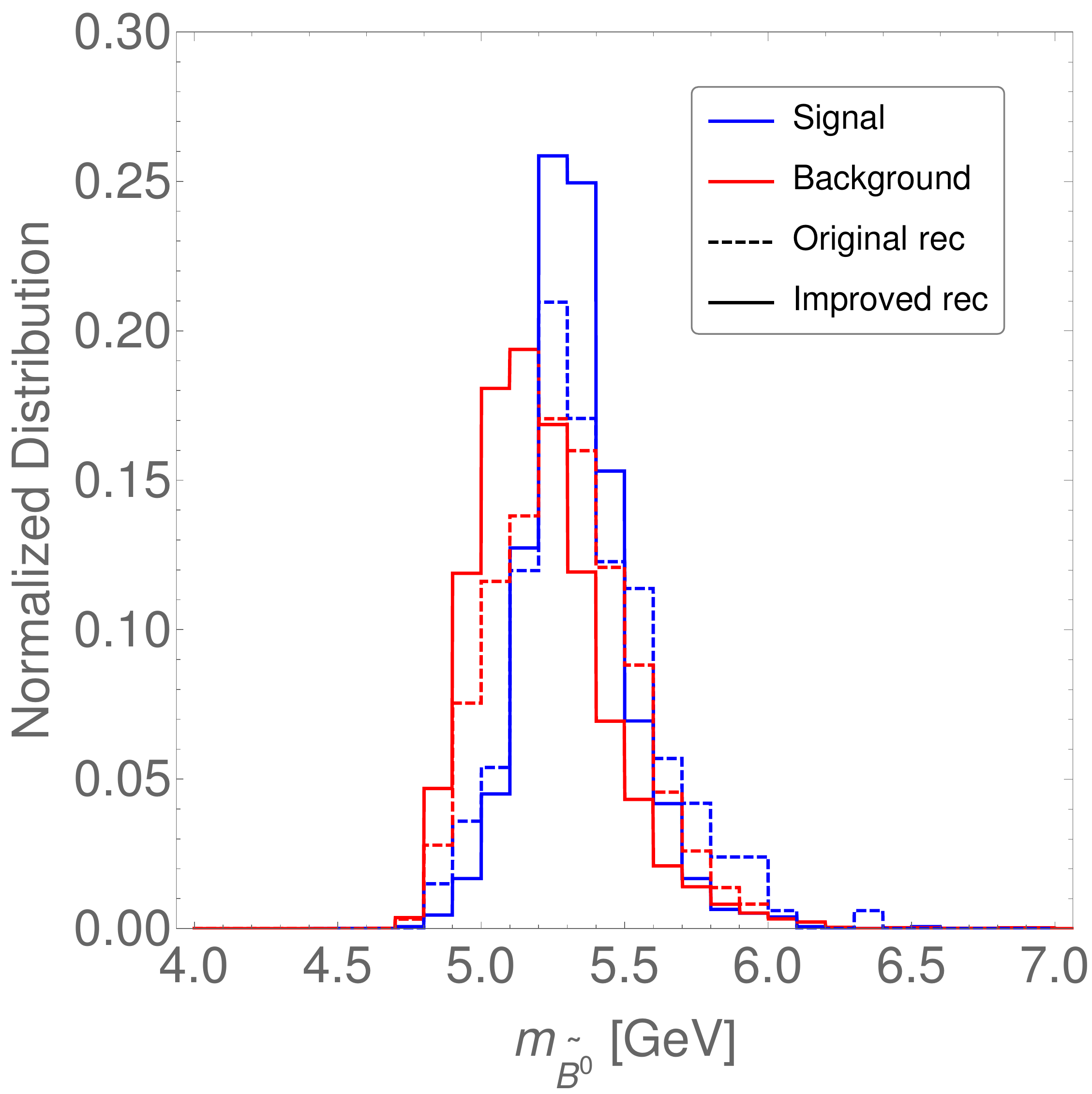}
\caption{Normalized distributions of the reconstructed $B$-meson mass for the $B^0\to K^{\ast 0} \tau^+ \tau^-$ measurement before cuts 1-3 are applied. Here the spatial resolution for the tracker is assumed to be 5$\mu$m (left) and 10$\mu$m (right). The dashed and solid curves are based on the reconstruction scheme developed for an ideal tracker (see Sec.~\ref{sec:pheno}) and the one improved for a tracker with finite spatial resolution (see footnote~\ref{lab:rec}).} 
\label{fig:micron}
\end{figure}

With the signal efficiencies and background rejections summarized in Tab.~\ref{tab:resultks}, we are able to calculate the sensitivities of measuring the four $b\to s \tau^+\tau^-$ benchmark channels at the future $Z$ factories. The outcomes are summarized in Tab.~\ref{tab:comparison} and also presented as a bar chart in Fig.~\ref{fig:res}~\footnote{In the FCC-ee~\cite{Kamenik:2017ghi,Abada:2019lih} and CEPC~\cite{CEPCStudyGroup:2018ghi} literatures, the $b\to s\tau^+\tau^-$ measurements at the $Z$ pole are qualitatively or semi-quantitatively considered to be promising. For the $B^0\to K^{\ast 0} \tau^+\tau^-$ channel, a crude signal yield of $\mathcal{O}(10^2)$ is predicted at Tera-$Z$. This result is compatible with our $B^0\to K^{\ast 0} \tau^+\tau^-$ preselection shown in Tab.~\ref{tab:resultks}. But, a systematic evaluation on its major backgrounds is missing and also no kinematic cuts (such as cut 1-3 and the $B$-meson mass window cut developed in this study) are explored. Consequently, no analysis-level prediction on the $B^0\to K^{\ast 0} \tau^+\tau^-$ measurement is made. We also notice that no quantitative predictions for the $B_s\to\phi\tau^+\tau^-$ and $B^+\to K^+\tau^+\tau^-$ channels are presented in these literatures. As for the $B_s\to \tau^+\tau^-$ channel, its sensitivity is roughly estimated to be $< 2\times 10^{-5}$ at Tera-$Z$, by scaling the sensitivity at Belle~II~\cite{CEPCStudyGroup:2018ghi}. The work pursued in this paper is thus the first dedicated sensitivity study on the $b\to s\tau^+\tau^-$ measurements at the future $Z$ factories, to our knowledge.}. The projected sensitivities of Belle~II~\cite{Kou:2018nap} and the upgraded LHCb~\cite{Bediaga:2018lhg}, if being available to us, are also shown as a reference. The future $Z$ factories demonstrate great potential for these measurements. For the $B^+ \to K^+ \tau^+ \tau^-$ and $B_s \to \tau^+\tau^-$ channels, Tera-$Z$ improve their sensitivities at Belle~II or LHCb by one to two orders. As for the other two, no sensitivities have been predicted by Belle~II or LHCb to our knowledge. On top of that, $10 \times$Tera-$Z$ makes a further improvement to the sensitivities by a factor $\sim \sqrt{10}$. 

Despite these encouraging outcomes, there exist several issues worth our noting and further thinking. First, to make these measurements realistic, we need to conduct a more solid study on the $B$-meson decays mediated by the $b\to c \bar c s  (\bar u d, \tau \nu)$ transitions. Such $B$-meson decays comprise the major backgrounds for these measurements, as discussed above. As shown in Tab.~\ref{tab:comparison} and Fig.~\ref{fig:res}, varying the experimentally measured backgrounds by one sigma and the semi-quantitatively estimated ones by a factor of two will cause a shift $\sim 25-40\%$ to the precisions of measuring $B^0\to K^{\ast 0} \tau^+ \tau^-$, $B_s\to\phi \tau^+ \tau^-$ and $B^+ \to K^+ \tau^+ \tau^-$, and a shift $\sim 15-20\%$ to that of measuring $B_s \to \tau^+ \tau^-$. The future $Z$ factories may improve our knowledge on the $b \to c$ physics simultaneously and significantly. Actually, some of these measurements, such as $b\to c \tau \nu$, can serve as a LFU probe also, as discussed in Sec.~\ref{sec:intro}.    

Second, a tracker with high spatial resolution is important for the suggested $b\to s\tau^+\tau^-$ measurements. Given that the signal $B$ mesons can be fully reconstructed based on the tracker in three of the four benchmark measurements, the tracker resolution essentially sets up a threshold for their expected sensitivities. As shown in Fig.~\ref{fig:mtB}, the invariant mass of $B$ mesons can be reconstructed at a level $\lesssim \mathcal O(0.1)$GeV (more precisely, with an FWHM $\sim 0.05$GeV for its distributions) for the measurements except $B_s \to \tau^+\tau^-$, with an ideal tracker. To demonstrate the impacts of the finite tracker resolution, we vary its value from the ideal case to a level of 5$\mu$m and 10$\mu$m, and show the normalized distributions of the reconstructed $B$-meson mass for the $B^0\to K^{\ast 0} \tau^+ \tau^-$ measurement in Fig.~\ref{fig:micron}. Comparing with the panels in the 1st row of Fig.~\ref{fig:mtB}, we find that the FWHM of the signal distributions are broadened by a factor $\sim 8 -10$ for the original reconstruction scheme (dashed curves) and a factor $\sim 6 - 8$ for the improved one (solid curves)~\footnote{\label{lab:rec}There is no template available for simulating the spatial resolution of the tracker in Delphes3~\cite{deFavereau:2013fsa}. In this study, we smear each vertex using a 3D Gaussian template, with its variance being 5$\mu$m and 10$\mu$m. We then introduce one vector variable for each vertex to compensate for this smearing in the $B$-meson reconstruction. The reconstruction is not sensitive to the shift along the vertex displacement direction, according to Eq.~(\ref{eq:fullkinematics1}), so we turn on the transverse components of each vector variable only. Totally we introduce six, six, six and four new parameters for the reconstructions of $B^0\to K^{\ast 0} \tau^+ \tau^-$, $B_s\to\phi \tau^+ \tau^-$, $B^+ \to K^+ \tau^+ \tau^-$ and $B_s \to \tau^+ \tau^-$, respectively. The scalar sum of these vector variables in each event is required to be less than three times of the tracker resolution. These new parameters, together with the ones introduced in Sec.~\ref{sec:pheno}, are eventually determined by minimizing the corespondent reconstruction error. For illustration, we present the distributions of the reconstructed $B$-meson mass based on the original scheme and this improved one for the $B^0\to K^{\ast 0} \tau^+ \tau^-$ measurement in Fig.~\ref{fig:micron}.}. Such a smearing effect necessarily reduces the expected sensitivities of these measurements. With the improved reconstruction scheme for the signal $B$ mesons, we implement cuts 1-3 introduced above for these measurements and define their  sensitivities in the modified $m_{\tilde B}$ windows, namely [5.2,5.8]GeV, [5.4,6.0]GeV, [5.2,5.8]GeV and [5.0,6.0]GeV. As summarized in Tab.~\ref{tab:resultks} and Fig.~\ref{fig:res}, we find the expected sensitivities to be reduced by a factor $\sim 3-6$ for the measurements except $B_s \to \tau^+\tau^-$, compared to the case with an ideal tracker. This outcome may serve as the inputs for setting up the baseline parameters of the tracker at the future $Z$ factories.   
   
At last, Tera-$Z$ fails to reach a sensitivity level required for measuring these $b\to s\tau^+\tau^-$ channels in the SM (see Tab.~\ref{tab:channels}), even with an ideal tracker, despite its great potential to explore their underlying new physics (see Subsec.~\ref{ssec:eft}). $10 \times$Tera-$Z$ improves this situation to some extent. With an ideal tracker, it can measure the SM $B^0\to K^{\ast 0} \tau^+ \tau^-$ and $B^+ \to K^+ \tau^+ \tau^-$ directly and yields an sensitivity not far from measuring the SM $B_s\to\phi \tau^+ \tau^-$. This fact could serve as a reference for a later optimization of the operation scenarios at the $Z$ pole for both CEPC and FCC-ee.

\subsection{Interpretation in EFT}
\label{ssec:eft}

As an example of physical interpretation, in this subsection we will project the expected sensitivities of measuring the $b\to s\tau^+\tau^-$ benchmark channels at the future $Z$ factories to the EFT. The uncertainties arising from the background evaluation will not be considered. For simplicity, we only include 6D (pseudo)vector operators of $b\to s\tau^+\tau^-$ in the EFT, following the discussions in~\cite{Kamenik:2017ghi}, and neglect scalar and tensor ones. Some of the latter operators have been tightly constrained by data~\cite{Bobeth:2011st}. Explicitly, this EFT Lagrangian is given by 
\begin{equation}
\mathcal{L}^{\rm eff}_{b\to s\tau^+\tau^-} = \mathcal{L}^{\rm SM}_{b\to s\tau^+\tau^-}+ \frac{4 G_F V_{tb}V^\ast_{ts}}{\sqrt{2}}(\delta C^\tau_9 O_9^\tau+\delta C^\tau_{10} O_{10}^\tau+C_{9}^{\prime\tau} O_{9}^{\prime\tau}+ C_{10}^{\prime\tau}  O_{10}^{\prime\tau}) +\rm{h.c.}~,
\label{eq:NCH}
\end{equation}
with 
\begin{equation}
O_{9(10)}^{\tau}=\frac{\alpha}{4\pi}[\bar{s}\gamma^\mu P_L b ][\bar{\tau}\gamma_\mu(\gamma^5)\tau]~,~O_{9(10)}^{\prime\tau}=\frac{\alpha}{4\pi}[\bar{s}\gamma^\mu P_R b ][\bar{\tau}\gamma_\mu(\gamma^5)\tau]~.
\label{eq:NCO910}
\end{equation}
These operators are well-motivated in new physics. They can be generated by either colorless or colored spin-one flavor mediators. The simplest colorless example might be family non-universal $Z'$ boson~\cite{Langacker:2000ju,Barger:2009eq,Barger:2009qs}, where the said operators arise from $s$ channel. This case can be extended to the context with an extra SU(2) gauge triplet~\cite{Boucenna:2016qad,Chiang:2017hlj,Kumar:2018kmr,Asadi:2018wea,Greljo:2018ogz,Abdullah:2018ets,Greljo:2018tzh,Gomez:2019xfw}. It has been shown that in this setup the $b\to s\ell^+\ell^-$ and $b\to c\tau\nu$ anomalies can be resolved by the flavor mediators of $Z'$ and $W'$, respectively~\cite{Boucenna:2016qad,Chiang:2017hlj,Kumar:2018kmr,Asadi:2018wea,Greljo:2018ogz,Abdullah:2018ets,Greljo:2018tzh,Gomez:2019xfw}. The colored example is leptoquarks where the said operators arise from $t$ channel. Currently, the electroweak-singlet vector leptoquark, often referred to as leptoquark $U_1$, is favored by global fit of data~\cite{Barbieri:2016las,Barbieri:2017tuq,Kumar:2018kmr}. Notably, electroweak-loop diagrams in the SM can also contribute to the $O_{9(10)}^{\tau}$ operators, yielding $C^\tau_{9(10)}|_{\rm SM} \approx 4.07(-4.31)$ at $\mu_b=4.8$GeV~\cite{DescotesGenon:2011yn}. The effect has been included as part of $\mathcal{L}_{b\to s\tau^+\tau^-}^{\rm SM}$ in this Lagrangian. 

This Lagrangian predicts the branching ratios, in terms of Wilson coefficients, to be 
\begin{eqnarray}
\label{eq:Ksoperator}
\text{BR}(B^0\to K^{\ast 0} \tau^+ \tau^-)\times 10^{7} &=& 0.98+0.38 \delta C^\tau_9  -0.14 \delta C^\tau_{10} -0.30 C_{9}^{\prime\tau} +0.12 C_{10}^{\prime\tau}  \nonumber \\
&&  -0.08 \delta C^\tau_9 C_{9}^{\prime\tau}  -  0.03 \delta C^\tau_{10}C_{10}^{\prime\tau}  + 0.05 (\delta C^\tau_9 ) ^2 \nonumber \\ &&  +0.02 (\delta C^\tau_{10}) ^2  +0.05 (C_{9}^{\prime\tau} ) ^2+ 0.02 (C_{10}^{\prime\tau} )^2~, \\ 
\text{BR}(B_s\to\phi \tau^+ \tau^-)\times 10^{7} &=& 0.86+0.34 \delta C^\tau_9  -0.11 \delta C^\tau_{10} -0.28 C_{9}^{\prime\tau} +0.10 C_{10}^{\prime\tau}   \nonumber \\
&&  -0.08 \delta C^\tau_9 C_{9}^{\prime\tau} -  0.02 \delta C^\tau_{10}C_{10}^{\prime\tau}  + 0.05 (\delta C^\tau_9 ) ^2   \nonumber \\
&&
+0.01 (\delta C^\tau_{10}) ^2 +0.05 (C_{9}^{\prime\tau} ) ^2+ 0.01 (C_{10}^{\prime\tau} )^2~, \\
\text{BR}(B^+ \to K^+ \tau^+ \tau^-  )\times 10^{7} &=& 1.20+0.15 \delta C^\tau_9  -0.42 \delta C^\tau_{10} +0.15 C_{9}^{\prime\tau} -0.42 C_{10}^{\prime\tau}   \nonumber \\
&& +0.04 \delta C^\tau_9 C_{9}^{\prime\tau}  +  0.10 \delta C^\tau_{10}C_{10}^{\prime\tau}  + 0.02 (\delta C^\tau_9 ) ^2 \nonumber \\
&&
+0.05 (\delta C^\tau_{10}) ^2 +0.02 (C_{9}^{\prime\tau} ) ^2+ 0.05 (C_{10}^{\prime\tau} )^2~,
\end{eqnarray}
for the first three $b \to s \tau^+\tau^-$ channels~\cite{Capdevila:2017iqn}, and 
\begin{eqnarray}
\text{BR}(B_s \to \tau^+ \tau^-)\times 10^{7} &=&  7.7 - 1.8\delta C^\tau_{10} + 0.11 (\delta C^\tau_{10}) ^2 -1.8C_{10}^{\prime\tau}  \nonumber \\
&& + 0.11 (C_{10}^{\prime\tau} ) ^2 -0.22 \delta C^\tau_{10}C_{10}^{\prime\tau} ~,
\label{eq:ditoperator}
\end{eqnarray}
for $B_s\to \tau^+\tau^-$~\cite{Bobeth:2011st,Bobeth:2014tza}. Here the branching ratios are calculated for the $q^2$ range shown in Tab.~\ref{tab:channels}. All Wilson coefficients are defined at $\mu_b=4.8$GeV and assumed to be real.

\begin{figure}[h!]
\centering
\includegraphics[height=10cm]{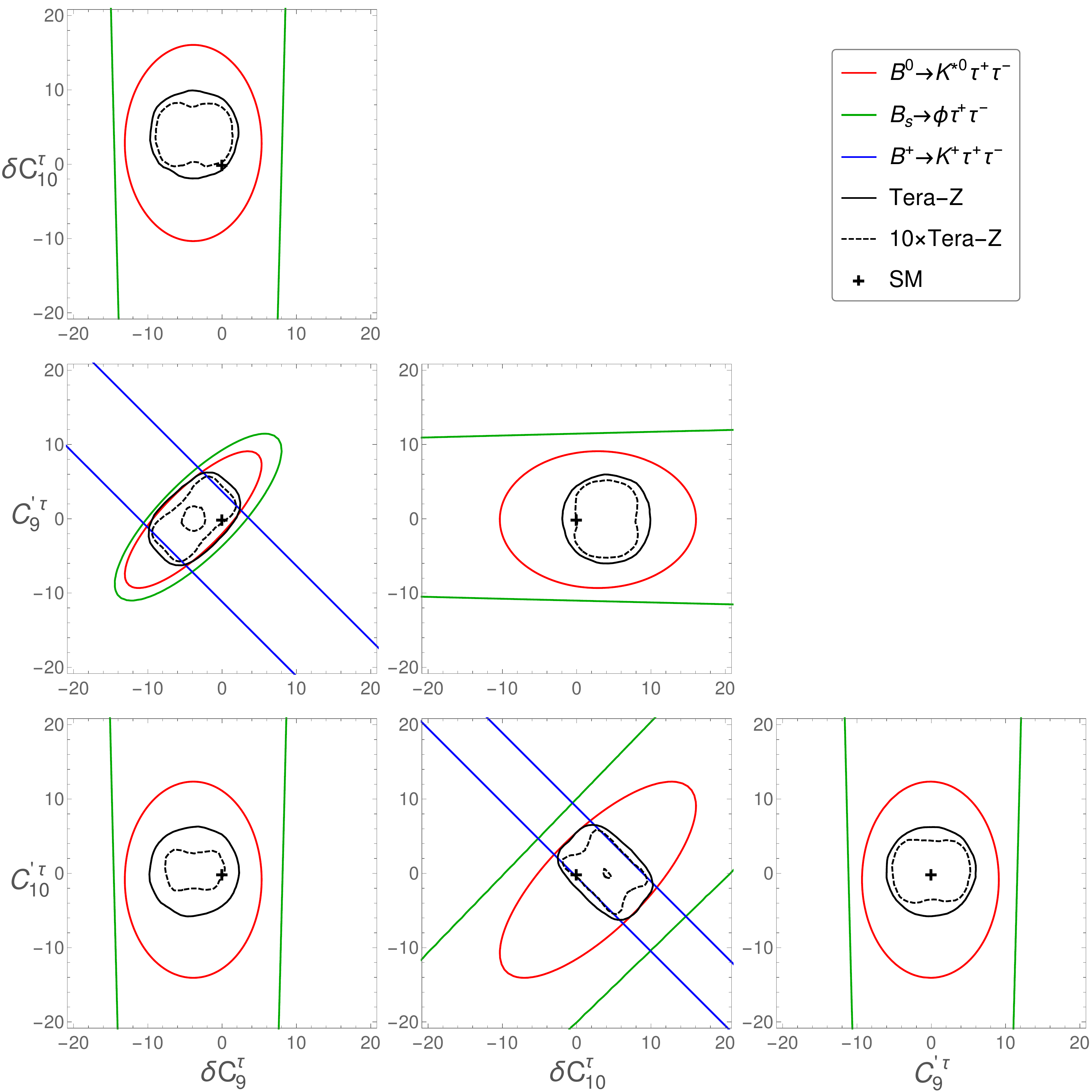}

\caption{2D marginalized constraints ($@ 1\sigma$ C.L.) for $\delta C^\tau_{9,10}$ and $C'^\tau_{9,10}$ at the future $Z$ factories, with an ideal tracker. The coordinate axes are in the unit of $\frac{4 G_F V_{tb}V^\ast_{ts}}{\sqrt{2}} \sim (0.86{\rm TeV})^{-2}$.}
\label{fig:2DM}
\end{figure}

\begin{figure}[h!]
\centering
\includegraphics[height=6cm]{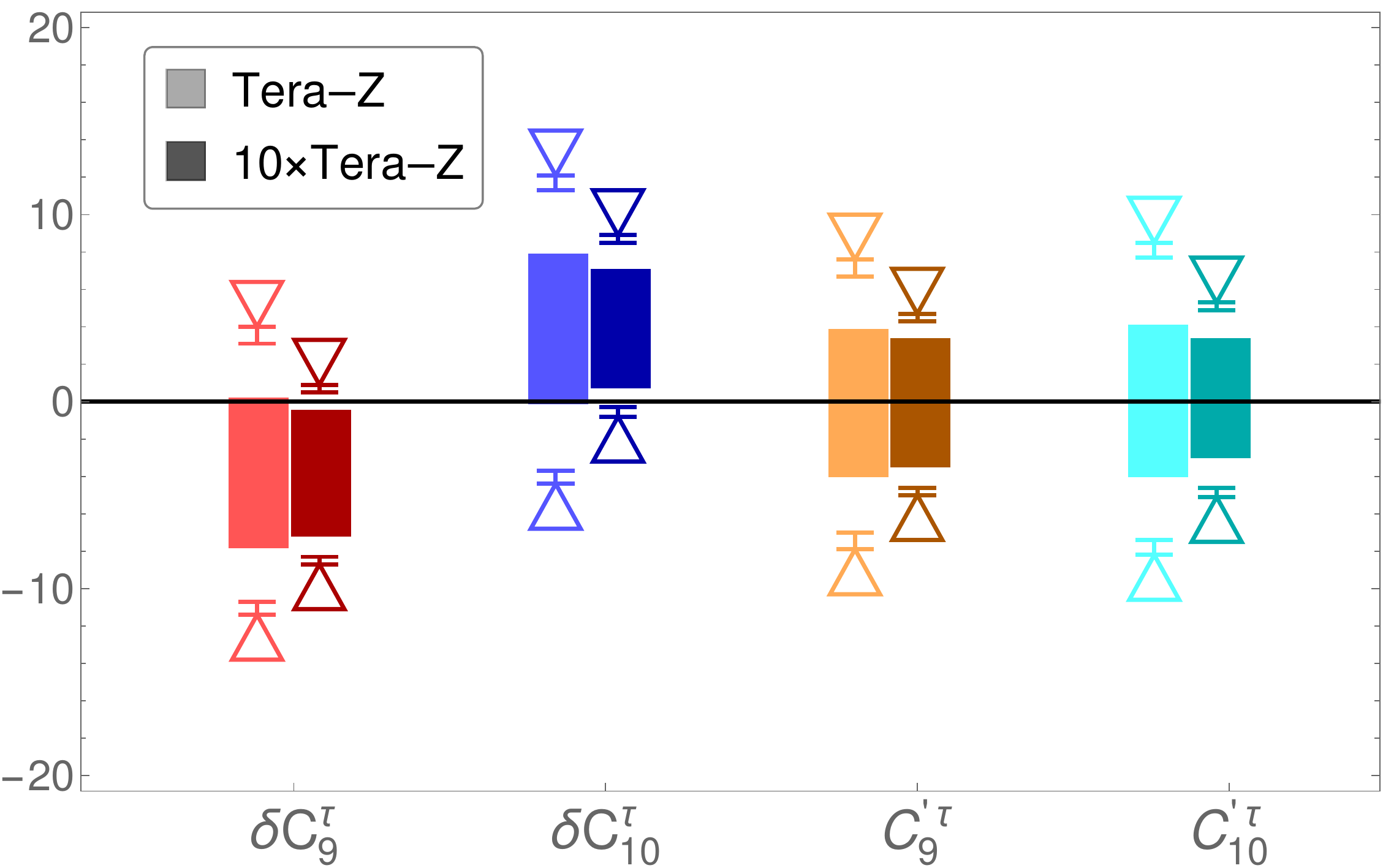}
\caption{1D marginalized constraints ($@ 1\sigma$ C.L.) for $\delta C^\tau_{9,10}$ and $C'^\tau_{9,10}$ at the future $Z$ factories. The double bars below the inverted triangle and above the triangle denote the sensitivities with a finite spatial resolution, $i.e.$, 5$\mu$m and 10$\mu$m respectively, for the tracker. The coordinate axes are in the unit of $\frac{4 G_F V_{tb}V^\ast_{ts}}{\sqrt{2}} \sim (0.86{\rm TeV})^{-2}$.
}
\label{fig:1DM}
\end{figure}

We present in Fig.~\ref{fig:2DM} the 2D marginalized constraints ($@ 1\sigma$ C.L.) for $\delta C^\tau_{9,10}$ and $C'^\tau_{9,10}$ at the future $Z$ factories. These contours are mainly determined by $B^0\to K^{\ast 0} \tau^+ \tau^-$, $B_s\to\phi \tau^+ \tau^-$ and $B^+ \to K^+ \tau^+ \tau^-$. Thereinto, the amplitudes of $B^0\to K^{\ast 0} \tau^+ \tau^-$ and $B_s\to\phi \tau^+ \tau^-$ are both calculated relying on the transition from $B$ meson to vector meson, namely $K^{\ast 0}$ or $\phi$~\cite{Kamenik:2017ghi}, except a difference caused by meson mass and decay form factor. This necessarily results in similar dependences of their branching ratios on $\delta C^\tau_{9,10}$ and $C'^\tau_{9,10}$ and hence close contour patterns for their constraints on the Wilson coefficients, as shown in  Eq.~(\ref{eq:Ksoperator}) and Fig.~\ref{fig:2DM}, respectively. The sensitivity of measuring $B_s\to \phi \tau^+\tau^-$ is relatively weak compared to that of measuring $B^0\to K^{\ast 0} \tau^+ \tau^-$, due to a lower production rate of $B_s$ mesons. The contours generated by this channel are thus bigger. Interestingly, $B^+ \to K^+ \tau^+ \tau^-$ demonstrates a strong complementarity to these two channels. According to Eq.~(\ref{eq:Ksoperator}), it generates two approximate degenerate directions along $\delta C_{9,10}^\tau \sim - C_{9,10}^{'\tau}$, while $B^0\to K^{\ast 0} \tau^+ \tau^-$ and $B_s\to\phi \tau^+ \tau^-$ generate two along $\delta C_{9,10}^\tau \sim  C_{9,10}^{'\tau}$. These Wilson coefficients thus can be well-constrained in all directions. As for the $B_s \to \tau^+ \tau^-$ measurement, its sensitivity is at least one order lower than the others. So its impacts on these contours can be neglected. 

By taking further marginalization, we generate the 1D constraints for $\delta C^\tau_{9,10}$ and $C^{\prime\tau}_{9,10}$ and show them in Fig.~\ref{fig:1DM}.     
Relying on the tracker spatial resolution, the constraints at Tera-$Z$ vary from below to above ten, while $10\times$Tera-$Z$ can constrain their magnitudes to universally below ten. This pushes the scale of new physics to probe from $\sim \mathcal O(1)$TeV to $\sim \mathcal O(10)$TeV. To make more sense of these numbers, let us consider the EFT discussed in~\cite{Capdevila:2017iqn} where electroweak symmetry is preserved. As suggested, explaining the $b\to c\tau\nu$ anomalies in this context requires $\delta C_{10}^\tau -\delta C_9^\tau \sim \mathcal O(10^2)$. The current measurements of $B^+\to K^+\tau^+\tau^-$ and $B_s\to\tau^+\tau^-$~\cite{TheBaBar:2016xwe,Aaij:2017xqt} probe these Wilson coefficients at a level $\sim\mathcal{O}(10^3)$ only. But, this scenario can be either confirmed or rejected in an unambiguous manner at future $Z$ factories, according to Fig.~\ref{fig:1DM}.

\section{Summary and Outlook}
\label{sec:conclusion}

The measurement of flavor physics is one of the most important tasks for the $Z$ factories at the next-generation $e^-e^+$ colliders. In this paper we pursued a dedicated sensitivity study on the $b\to s \tau^+\tau^-$ measurements at the future $Z$ factories. This is highly motivated for addressing the LFU-violating puzzles such as $R_{K^{(\ast)}}$, $R_{D^{(*)}}$ and $R_{J/\psi}$ anomalies, and in a more general context, for testing LFU as a fundamental rule in particle physics. The advantages of the future $Z$ factories for achieving this task are two-fold. First, the planned Tera-$Z$ for CEPC and its upgraded version of $10\times$Tera-$Z$ at FCC-ee will both generate a number of clean $b$-hadron events. Explicitly, the productions of $B^0/\bar{B^0}$ and $B^\pm$ at Tera-$Z$ are comparable to those at Belle~II, while the $B_s$/$\bar{B}_s$ productions are nearly two orders more. Second, the produced $b$ hadrons at the $Z$ pole are highly boosted. These $b$ hadrons tend to decay with a larger displacement, with a cluster of energetic charged decay products, such that the relevant kinematics can be well measured in the tracker. The reconstruction of these $b$ hadrons is thus expected to be highly efficient. 

Explicitly, we conducted this study in the four benchmark channels: $B^0\to K^{\ast 0} \tau^+ \tau^-$, $B_s\to\phi \tau^+ \tau^-$, $B^+ \to K^+ \tau^+ \tau^-  $ and $B_s \to \tau^+ \tau^-$. These $b\to s\tau^+\tau^-$ events are not visible to the detector completely, due to the generation of neutrinos in the $\tau$ decays. To address this difficulty, we develop a scheme to reconstruct the signal $B$ mesons that works for all of the four $b\to s\tau^+\tau^-$ channels, if both $\tau$ lepton decay to $\pi^\pm\pi^\pm\pi^\mp\nu$. In such a scheme, the $B$ mesons are reconstructed based on two arguments:  the momentum of the decaying particle should be collimated with its displacement and the $\tau$ leptons should be both on-shell. This scheme is fully tracker-based and makes use of the advantages said above. Eventually it yields a full reconstruction of the signal events except $B_s \to \tau^+ \tau^-$.

The major backgrounds for these measurements are expected to arise from the Cabibbo-favored $b\to c+X$ processes, namely $b\to c\bar{c}s$, $b\to c\tau \nu$ and $b\to c\bar{u}d$, where either one or both $\tau$ leptons in the signal are faked by the charged $D$ meson(s). As is well-known, $D_s^\pm$ and $D^\pm$ mesons have a mass and lifetime comparable to those of $\tau$ leptons in the SM and they can also decay to a vertex of $\pi^\pm\pi^\pm\pi^\mp$ with extra particles. Yet, most of these backgrounds have not been experimentally measured or theoretically calculated. The information needed for the sensitivity analysis is hence largely missing. We introduced a semi-quantitative strategy to address this difficulty. Explicitly, we first introduced a well-measured decay mode of $B$ mesons as a  reference, then estimated the suppression factor of the given background relative to this reference, according to the effects reviewed in~\cite{Bevan:2014iga}, and at last calculated this background by scaling. The signal and backgrounds were simulated by decaying $B$ mesons exclusively and the intermediate particles of their decays ($\tau$ leptons, $D$ mesons, etc.) inclusively. With this design, we are able to evaluate the $Z$-factory capability for achieving this task quantitatively and meanwhile are ready to adapt the expected sensitivities obtained in this study to the future updates on the major backgrounds whenever they become available.

The analysis results at Giga-$Z$, Tera-$Z$ and $10\times$Tera-$Z$, with the tracker resolution varying from an ideal case to $5\mu$m and $10\mu$m, were summarized in Tab.~\ref{tab:resultks} and Fig.~\ref{fig:res}. Their interpretations in the EFT were also presented in Sec.~\ref{ssec:eft}. The outcomes demonstrate a great potential of the future $Z$ factories in measuring the $b\to s \tau^+\tau^-$ transitions and exploring its underlying physics. For example, compared to the Belle~II ones, the expected precisions of measuring $B^+ \to K^+ \tau^+ \tau^-$ are one to two orders higher at Tera-$Z$, resting upon the tracker resolution, and at least two orders higher at $10\times$Tera-$Z$. Accordingly, the new-physics scale expected to probe is pushed from $\sim 1$TeV to $\sim10$TeV. Despite this, Tera-$Z$ fails to reach a sensitivity level to measure these benchmark channels in the SM, even with an ideal tracker. $10 \times$Tera-$Z$ improves this situation to some extent, yielding a marginal sensitivity to them except $B_s\to\phi \tau^+ \tau^-$, with an ideal tracker. These outcomes may serve as the inputs for setting up the detector baseline parameters of the future $Z$ factories and as a reference for a later optimization of their operation scenarios at the $Z$ pole.  

The sensitivities demonstrated in this study could be further improved. For example, the reconstruction scheme of $B$ mesons developed in this study is fully tracker-based. It has been shown that the decay of $\pi^0\to\gamma\gamma$ can be well-reconstructed with high-granularity calorimeters at the future lepton colliders~\cite{Shen:2019yhf}. For that case, the message from the calorimeters could be combined for improving the $B$-meson reconstruction, since both $\tau$ leptons and $D$ mesons can decay into three tracks with extra pions. More than that, advanced data-mining tools such as deep neural network (DNN) may improve the expected sensitivities at a significant level. For the precision machines such as CEPC, FCC-ee and ILC, the primary Higgs and electroweak processes at low beam energy are dominated by hadronic modes.  As demonstrated in~\cite{Li:2020vav}, the tool of DNN is powerful in synergizing the kinematic messages at hadron level, and hence may significantly improve many of the baseline precisions presented in the literatures. At last, we have strong motivation to extend this study to the $b\to c\tau \nu$ measurements at the future $Z$ factories. As discussed in Sec.~\ref{sec:intro}, the $b\to c\tau \nu$ transitions are mediated by FCCC, and hence complement the FCNC-meditated $b\to s \tau^+\tau^-$ transitions. To fully test LFU or to generate a global picture on LFU, we need to take both of them as inputs (for a recent study on the measurement of $B_c \to \tau \nu$ at CEPC, see~\cite{Zheng:2020emi}). Analysis results based on these thinkings will be presented in our future work.

\section*{Acknowledgements}
We thank Elisabetta Barberio, Lorenzo Calibbi, Franco Grancagnolo, Seung J. Lee, Manqi Ruan, and Lian-Tao Wang for useful discussions. This research was supported partly by the General Research Fund (GRF) under Grant No 16302117 and partly by the Area of Excellence under the Grant No AoE/P-404/18-3. Both grants were issued by the Research Grants Council of Hong Kong S.A.R..

\appendix
\section{Detailed Cut Flows and Tera-$Z$ Yields for the $b\to s \tau^+\tau^-$ Measurements}
\label{app:cutflow}

\begin{table}[h!]
\centering
\begin{tiny}
\begin{tabular}{cccccccc} 
\hline
Channel    & BR   & $\epsilon_{\rm pre}$  & $\epsilon_{1}$  & $\epsilon_{2}$  & $\epsilon_{3}$  & $\epsilon_{m_{\tilde B}}$  & Tera-$Z$ Yield       \\ 
\hline
 $B^0\to K^{\ast 0} \tau^+\tau^-$                 & $9.8\times 10^{-8}$   & $3.3\times 10^{-3}$  & $6.3\times 10^{-1}$ & $7.3\times 10^{-1}$ & $8.8\times 10^{-1}$   & $6.4\times 10^{-1}$ & $1.0\times 10^{1}$     \\ 
\hline
 $B^0\to K^{\ast 0} D_s^{(\ast)-}\tau^+ \nu$  & $3.0\times 10^{-5}$  & $4.9\times 10^{-4}$ & $1.9\times 10^{-1}$  & $5.0\times 10^{-1}$ & $9.4\times 10^{-1}$ & $5.4\times 10^{-2}$ & $8.7$                \\
$B_s\to \bar{K}^{\ast 0} D^{(\ast)-}\tau^+\nu$      & $4.6\times 10^{-4}$  & $1.9\times 10^{-4}$ & $1.7\times 10^{-1}$  & $4.9\times 10^{-1}$ & $5.2\times 10^{-1}$ & $1.0\times 10^{-1}$  & $1.2\times 10^1$                \\
$B_s\to K^{\ast 0} D^{(\ast)+} D_s^{(\ast)-}$ & $1.2\times 10^{-2}$  & $4.2\times 10^{-4}$ & $3.6\times 10^{-2}$ & $4.1\times 10^{-1}$ & $4.5\times 10^{-1}$ & $8.8\times 10^{-2}$ & $9.3\times 10^1$   \\
$B_s\to \bar{K}^{\ast 0} D_s^{(\ast)+} D^{(\ast)-}$ & $1.2\times 10^{-2}$  & $4.2\times 10^{-4}$ & $3.6\times 10^{-2}$ & $4.1\times 10^{-1}$ & $4.5\times 10^{-1}$ & $8.8\times 10^{-2}$ & $9.3\times 10^1$   \\
$B^0\to K^{\ast 0} D^{(\ast)+} D^{(\ast)-}$   & $1.2\times 10^{-2}$  & $1.6\times 10^{-4}$ & $3.3\times 10^{-2}$ & $3.8\times 10^{-1}$ & $2.7\times 10^{-1}$ & $<2.3\times 10^{-2}$ & $<1.8\times 10^1$   \\
$B^0\to K^{\ast 0} D_s^{(\ast)+} D_s^{(\ast)-}$& $1.6\times 10^{-3}$ & $7.3\times 10^{-4}$ & $7.1\times 10^{-2}$ & $4.0\times 10^{-1}$ & $6.4\times 10^{-1}$ & $1.5\times 10^{-2}$ & $4.0\times 10^1$   \\
\hline
\end{tabular}
\end{tiny}
\caption{Detailed cut flows and Tera-$Z$ yields for the $B^0\to K^{\ast 0}\tau^+\tau^-$ measurement. $\epsilon$ denotes the efficiency for each cut defined in the text.}
\end{table}

\begin{table}[h!]
\centering
\begin{tiny}

\begin{tabular}{cccccccc} 
\hline
Channel    & BR   & $\epsilon_{\rm pre}$  & $\epsilon_{1}$  & $\epsilon_{2}$  & $\epsilon_{3}$  & $\epsilon_{m_{\tilde B}}$  & Tera-$Z$ Yield       \\ 
\hline 
$B_s\to\phi \tau^+ \tau^-$ & $8.6\times 10^{-8}$                  & $2.0\times 10^{-3}$ & $6.3\times 10^{-1}$ & $7.3\times 10^{-1}$ & $8.8\times 10^{-1}$ & $6.3\times 10^{-1}$ & $1.4$ \\ 
\hline
$B_s\to \phi D_s^{(\ast)-}\tau^+ \nu$ & $2.6\times 10^{-5}$       & $4.3\times 10^{-4}$ & $2.5\times 10^{-1}$ & $5.4\times 10^{-1}$ & $8.1\times 10^{-1}$ & $1.2\times 10^{-1}$ & $4.6$ \\ 
$B^0 \to \phi D^{(\ast)-} \tau^+ \nu$ & $4.0\times 10^{-7}$       & $2.2\times 10^{-4}$ & $1.6\times 10^{-1}$ & $6.3\times 10^{-1}$ & $6.7\times 10^{-1}$ & $8.6\times 10^{-2}$ & $0.1$  \\ 
$B_s \to\phi  D_s^{(\ast)+} D_s^{(\ast)-} $ & $1.6\times 10^{-3}$ & $7.0\times 10^{-4}$ & $8.3\times 10^{-2}$ & $3.1\times 10^{-1}$ & $6.0\times 10^{-1}$ & $2.9\times 10^{-2}$ & $1.6\times 10^{1}$ \\ 
$B^0 \to \phi D_s^{(\ast)+} D^{(\ast)-}  $ & $1.6\times 10^{-3}$  & $2.7\times 10^{-4}$ & $6.1\times 10^{-2}$ & $3.8\times 10^{-1}$ & $4.4\times 10^{-1}$ & $7.7\times 10^{-1}$ & $4.1\times 10^{1}$ \\ 
$B_s \to \phi D^{(\ast)+} D^{(\ast)-}  $ & $3.2\times 10^{-3}$    & $1.4\times 10^{-4}$ & $2.1\times 10^{-2}$ & $4.0\times 10^{-1}$ & $5.5\times 10^{-1}$ & $<6.7\times 10^{-2}$  & $<3.2$ \\ 
\hline 
\end{tabular} 
\caption{Detailed cut flows and Tera-$Z$ yields for the $B_s\to \phi\tau^+\tau^-$ measurement. $\epsilon$ denotes the efficiency for each cut defined in the text.}
\end{tiny}
\end{table}

\begin{table}[h!]
\centering
\begin{tiny}
\begin{tabular}{cccccccc} 
\hline
Channel    & BR   & $\epsilon_{\rm pre}$  & $\epsilon_{1}$  & $\epsilon_{2}$  & $\epsilon_{3}$  & $\epsilon_{m_{\tilde B}}$  & Tera-$Z$ Yield       \\ 
\hline 
$B^+ \to K^+ \tau^+ \tau^-  $           & $1.2\times 10^{-7}$  & $5.9\times 10^{-3}$  & $6.3\times 10^{-1}$ & $6.9\times 10^{-1}$ & $9.0\times 10^{-1}$   & $6.0\times 10^{-1}$ & $2.0\times 10^{1}$     \\ 
\hline
$B^+ \to K^+ D_s^{(\ast)-}  \tau^+ \nu$ & $9.5 \times 10^{-5}$   & $3.3\times 10^{-3}$  & $2.1\times 10^{-1}$ & $4.7\times 10^{-1}$ & $7.9\times 10^{-1}$   & $2.6\times 10^{-2}$ & $7.5\times 10^{1}$     \\ 
$B^+ \to K^+ D_s^{(\ast)+} D_s^{(\ast)-}$ & $1.6 \times 10^{-3}$ & $5.3\times 10^{-3}$  & $6.7\times 10^{-2}$ & $3.0\times 10^{-1}$ & $7.2\times 10^{-1}$   & $2.5\times 10^{-2}$ & $3.6\times 10^{2}$     \\ 
$B^+ \to K^+ D^{(\ast)+} D^{(\ast)-}$ &  $2.8 \times 10^{-3}$    & $5.8\times 10^{-4}$  & $3.7\times 10^{-2}$ & $4.9\times 10^{-1}$ & $5.0\times 10^{-1}$   & $5.4\times 10^{-2}$ & $8.7\times 10^{1}$     \\ 
\hline 
\end{tabular} 

\end{tiny}
\caption{Detailed cut flows and Tera-$Z$ yields for the $B^+\to K^+\tau^+\tau^-$ measurement. $\epsilon$ denotes the efficiency for each cut defined in the text.}
\end{table}

\begin{table}[h!]
\centering
\begin{tiny}
\begin{tabular}{cccccccc} 
\hline
Channel    & BR  & $\epsilon_{\rm pre}$  & $\epsilon_{1}$  & $\epsilon_{2}$  & $\epsilon_{3}$  & $\epsilon_{m_{\tilde B}}$  & Tera-$Z$ Yield       \\ 
\hline 
$B_s \to \tau^+ \tau^-$ & $7.7\times 10^{-7}$                       & $9.2\times 10^{-3}$  & $5.4\times 10^{-1}$ & $6.4\times 10^{-1}$ & $6.7\times 10^{-1}$   & $6.6\times 10^{-1}$ & $3.4\times 10^{1}$  \\  
\hline
$B_s \to D_s^{(\ast)-} \tau^+\nu$ & $2.7\times 10^{-2}$             & $1.0\times 10^{-2}$  & $2.2\times 10^{-1}$ & $4.4\times 10^{-1}$ & $5.8\times 10^{-1}$   & $1.9\times 10^{-1}$ & $9.5\times 10^{4}$    \\
$B^0 \to D^{(\ast)-} \tau^+\nu$ & $2.7\times 10^{-2}$               & $4.1\times 10^{-3}$  & $2.0\times 10^{-1}$ & $5.4\times 10^{-1}$ & $4.5\times 10^{-1}$   & $1.4\times 10^{-1}$ & $9.3\times 10^{4}$    \\
$B_s \to D_s^{(\ast)+} D_s^{(\ast)-}$ & $3.2\times 10^{-2}$         & $1.2\times 10^{-2}$  & $5.2\times 10^{-2}$ & $3.2\times 10^{-1}$ & $5.4\times 10^{-1}$   & $4.2\times 10^{-1}$ & $4.7\times 10^{4}$    \\
$B^0 \to D_s^{(\ast)+} D^{(\ast)-}$ & $4.0\times 10^{-2}$           & $4.9\times 10^{-3}$  & $5.4\times 10^{-2}$ & $4.2\times 10^{-1}$ & $3.4\times 10^{-1}$   & $3.9\times 10^{-1}$ & $6.9\times 10^{4}$    \\
$B^0\to D^{(\ast)-} \pi^\pm\pi^\pm\pi^\mp $ & $3.9\times 10^{-3}$   & $1.7\times 10^{-3}$  & $1.2\times 10^{-1}$ & $2.8\times 10^{-1}$ & $1.8\times 10^{-1}$   & $3.4\times 10^{-1}$ & $1.6\times 10^{3}$    \\
$B^0\to D^{(\ast)-} a_1^+$ & $1.9\times 10^{-2}$                    & $9.3\times 10^{-3}$  & $1.9\times 10^{-1}$ & $5.2\times 10^{-1}$ & $1.1\times 10^{-1}$   & $5.1\times 10^{-1}$ & $1.2\times 10^{5}$    \\
$B_s\to D_s^{(\ast)-} \pi^\pm\pi^\pm\pi^\mp $ & $3.9\times 10^{-3}$ & $4.4\times 10^{-3}$  & $1.7\times 10^{-1}$ & $2.6\times 10^{-1}$ & $1.1\times 10^{-1}$   & $4.6\times 10^{-1}$ & $1.4\times 10^{3}$    \\
$B_s\to D_s^{(\ast)-} a_1^+$ & $1.9\times 10^{-2}$                  & $2.0\times 10^{-2}$  & $2.0\times 10^{-1}$ & $4.1\times 10^{-1}$ & $9.1\times 10^{-2}$   & $4.0\times 10^{-1}$ & $3.5\times 10^{4}$    \\
\hline 
\end{tabular} 
\end{tiny}
\caption{Detailed cut flows and Tera-$Z$ yields for the $B_s\to\tau^+\tau^-$ measurement. $\epsilon$ denotes the efficiency for each cut defined in the text.}
\end{table}

\newpage

\bibliography{CEPCb}
\end{document}